\documentclass{aa}  

\usepackage{graphicx}
\usepackage{txfonts}
\usepackage{lipsum}
\usepackage{subcaption}         
\usepackage{lscape}             
\usepackage{placeins}           

\usepackage{url}

\usepackage{mathrsfs}

\newcommand\XMM{XMM-Newton }

\begin{document}

   \title{The origin of X-ray intra-day variability in HBL PKS 2155-304}

   \author{W. Hu
          \inst{1,2},  J. L. Kang  \inst{2,3} ,   J. X. Wang  \inst{2,3},   G. C. Xiao \inst{1}  
          \and
          G. W. Ren  \inst{4} 
          }

   \institute{Department of Physics, Jinggangshan University, Ji'an 343009, P. R. China\\
              \email{huwen.3000@jgsu.edu.cn(WH)}; xiaogc@ihep.ac.cn(GCX)
           \and
             CAS Key Laboratory for Research in Galaxies and Cosmology, Department of Astronomy, University of Science and Technology of China, Hefei, Anhui 230026, P. R. China\\
             \email{jxw@ustc.edu.cn}
            \and
                School of Astronomy and Space Science, University of Science and Technology of China, Hefei 230026, P. R. China
        \and
        Department of Astronomy, Xiamen University, Xiamen, Fujian 361005, P. R. China\\
             }
   \date{Received XXXX; accepted XXXX}

 
  \abstract
   { 
The X-ray radiation from the high-frequency peaked BL Lac object (HBL) is believed to be produced by the energetic electrons accelerated in the inner regions of a relativistic jet. This is well beyond the current imaging capabilities of telescopes in any part of the electromagnetic spectrum. 
   The rapid X-ray variability can offer crucial information for understanding the nature of the most compact regions in the relativistic jets, 
   and the underlying particle acceleration processes at the dissipation distance. 
  }
   {
   The origin and physics of X-ray intra-day variability (IDV) in blazars, which is a long-standing issue, is studied by modelling the broad-band X-ray spectrum, the light curves (LCs), and the Fourier time lags.
}
   {We present the timing analysis of three archived \textit{\XMM} observations with a total exposure of $>80$ ks of PKS 2155-304, which is one of the brightest and most studied HBLs in the X-ray band.
   For each observation, we constructed averaged X-ray spectra in 0.5–10 keV band, as well as 100 s binned LCs in various  sub-bands.
    We performed the Bayesian power spectral density (PSD) analysis and Fourier time-lag analyses of the variable LCs.
    The results are carefully modelled in the context of a multi-zone jet model. 
   }
   { 
   PSD analysis reveals that the X-ray variability can be characterised by   red noise.
   The lag–frequency spectra measured in two observations show only the soft or negative lags, with the magnitude of the lags increasing as the frequency decreases.
    For another observation, the lag--frequency spectra are characterised by small positive or zero time lags at the lowest frequencies, which drops to negative values at higher frequencies. 
The magnitude of the soft lags ranges from $\sim5$ to $\sim40$ minutes,
and increases with the energy difference of two compared LCs. 
The observed X-ray spectra and lag-frequency spectra can both be successfully described by our proposed two-zone model,
with the physical parameters constrained in a fully acceptable space.
Moreover, the LC profiles at different energy bands can be satisfactorily reproduced by only varying the injection rate of the energetic electrons. 
 }
   {The IDV of PKS 2155-304 should be caused by the injection of energetic electrons, and accelerated by shocks formed in a weakly magnetised jet.}


   \maketitle
%

\section{Introduction}
Blazars are the most extreme class of active galactic nuclei (AGNs) with a powerful relativistic jet closely aligned to our line of sight \citep{Urry1995PASP,Netzer2015ARA&A}.
They consist of BL Lac objects and flat spectrum radio quasars (FSRQs).
The observed emission, which is characterised by non-thermal radiation, is dominated by Doppler-boosted radiation from relativistic jets. 
The spectral energy distribution (SED) of blazars typically exhibits a broad double-peaked structure in the $\nu F_\nu$ representation \citep[e.g.][]{Chen2018ApJS,Ghisellini2014Natur,Hu2024MNRAS}. 
The low-energy peak is commonly believed to be synchrotron radiation from relativistic electrons and positrons in the jet.
The physical mechanisms responsible for the high-energy peak is still under debate, and may involve either leptonic or hadronic scenarios.
According to the leptonic scenario, it is probably produced via the inverse-Compton (IC) upscattering off the relativistic electrons in the jets, 
with the target photon fields being either  the internal synchrotron photons or external photons, 
such as those from the surrounding accretion disc, the broad-line region, or the dust torus \citep[e.g.][]{Ghisellini2009MNRAS,Dermer2009ApJ,Madejski2016annurev}. 
The radiation from the hadrons includes $\pi^0$ decay, photo-pion processes, and their induced cascades, 
as well as synchrotron emission of highly energetic protons ($E\gtrsim 10^{19}$ eV) in jets \citep[e.g.][]{Bottcher2013ApJ,Cao2020PASJ,Gasparyan2022MNRAS,Stathopoulos2024A}.

Across the entire electromagnetic spectrum, blazars exhibit flux and spectral variability on different timescales 
ranging from a few minutes to several decades.
Variations in flux on timescales of a few minutes to less than a day are often known as intra-day variability \citep[IDV; e.g.][and references therein]{Gaur2010ApJ,Devanand2022ApJ,Zhongli2021ApJ,Noel2022ApJS,Pavana2022MNRAS,Bhatta2025ApJ}.
Variability on timescales from days to a few months is referred to as short-term variability (STV)
and the flux variations on longer timescales are termed long-term variability \citep[LTV; e.g.][]{Goyal2020MNRAS,Zhanghy2021ApJ}.

Flux variations on IDV timescales are the most puzzling and are still not yet well understood.
In particular, the IDV of blazars in X-ray bands may not only provide insights into the physics at the base of the relativistic jet, 
but also it may be directly related to the activity in the innermost area of the central engine.
X-ray IDV is usually explained in terms of instabilities in the jet, such as turbulence or magnetic reconnection \citep{Marscher2014ApJ,Calafut2015JApA,Zhang2021MNRAS,Xu2023ApJS}.
On the other hand, IDV variations may be attributed to instabilities on or above the accretion discs for FSRQs, particularly in lower luminosity states \citep{Chakrabarti1993ApJ,Mangalam1993ApJ}.

Currently, the most popular model of blazars is the stationary or time-dependent one-zone homogeneous jet model, 
in which the large amount of radiation is treated as being produced by a single fast-moving plasma cloud inside the jet,
although extended jet components are required to reproduce radio observations  \citep[e.g.][]{Blandford1979,Konigl1981,Potter2013MNRAS,Zamaninasab2014Natur}.
Moreover, the observed rapid variability is more readily explained in terms of the much shorter acceleration and cooling  timescales of relativistic leptons in the jets. 
However, it should be noted that these models are facing strong challenges posed by observations, for example orphan flares, a few minutes timescale flares, significant hard X-ray excess
and an intrinsic hard TeV spectrum ($\Gamma\lesssim$ 1.5--1.9) peaking above 2--10 TeV.
Various theoretical models have been proposed to account for the physical phenomena \citep[e.g.][]{Moderski2003A&A,Ghisellini2005A&A,Boutelier2008MNRAS,Vuillaume2018A&A,Aguilar2022MNRAS,WangPhysRevD,Tan2024MNRAS}. 
All of these may be directly related to the structures inside an inhomogeneous jet. 
However, a significant number of free parameters are usually difficult to constrain and show large degeneracy.

The (quasi-)simultaneous  multi-wavelength (MWL) SEDs modelling and variability studies are fundamentally crucial for distinguishing between models 
and understanding the extreme physical conditions in the different emission regions of relativistic jets. 
Such studies require accurate and contemporaneous observations with the necessary energy coverage and time resolution.

In particular, Fourier frequency-dependent time lags between two different bands can provide tight constraints on model parameters,
and can offer unique insights into the different timescales associated with particle acceleration and radiation \citep{Finke2014ApJ,Finke2015ApJ,Lewis2016}.
Very recently, we performed an integrated study of the X-ray spectrum and time lags of ten archived \textit{XMM-Newton} observations with an exposure of $>40$ ks for Mrk 421 \citep{Hu2024ApJ}.
We found that both the signs and the amounts of the lags are different in different epochs, and it seems that the presence of the positive or negative lags is irrelevant to the broad-band X-ray spectral shape.
Based on the modelling of broad-band SED and inter-band time lags in the Fourier and time domains, we found that shocks 
inside the jet may be more plausible for powering the emission and triggering the rapid X-ray variability.

For this work we conducted a comprehensive study focusing on the X-ray emission from PKS 2155-304 \citep[at z=0.116;][]{Falomo1993ApJ},
which is classified as a high-frequency peaked (HBL) with the high energy astronomical observatory 1 observations in X-rays \citep{Schwartz1979ApJ}, 
and as a TeV blazar with the Durham Mark 6 atmospheric Cherenkov telescope \citep{Chadwick1999ApJ}.
The nuclear emission in these objects indicates that the accretion disc is radiatively inefficient, and any disc flux would be swamped by the jet emission.
Thus, any detectable IDV is unlikely to arise from the disc.
The extreme spectral properties and variability of these objects in the X-ray band and at very high energy (VHE) represent the highest energy particles 
and most violent acceleration processes in their jets. 
For the source there are several particularly interesting aspects that pose serious issues for single-zone blazar models. 
This object exhibits fast TeV flares with doubling times as short as a few minutes \citep{Aharonian2007ApJ},  very large Compton dominance during the $\gamma$-ray flare \citep{Aharonian2009A&A,Abramowski2012A&A}, and a variable X-ray hard tail \citep{Madejski2016ApJ,Gaur2017ApJ}.
These findings may reveal a structured or stratified inhomogeneous jet. 
Very recently, the different polarisation patterns observed at optical and X-ray bands have provided strong evidence of the structured scenario for the blazar zone \citep{Kouch2024A&A}.
Another interesting aspect is that several authors reported a possible detection of quasi-periodic oscillations (QPOs) on IDV timescales in the X-ray band \citep{Lachowicz2009A&A,Gaur2010ApJ} and in the polarised optical emission \citep{Pekeur2016MNRAS}.
This may shed new light on the physical processes and their associated emission mechanisms.

Our aim is to explore the origin of X-ray IDV  
by simultaneously reproducing the broad-band X-ray SED and Fourier lag-frequency spectra between various sub-bands of X-rays, as well as the X-ray light curve (LC) profiles.
We used public archive data of PKS 2155-304 taken by the EPIC (the European Photon Imaging Camera) instrument on board the XMM–Newton satellite,
which is uniquely well suited for monitoring and studying sub-hour variability.

The paper is structured as follows. In Sect. \ref{sec:data} we provide a brief description of the \textit{\XMM} observations of PKS 2155-304 and the data reduction method.
Fourier analysis and the results are presented in Sect. \ref{sec:method}. 
In Sect.  \ref{sec:model} we provide a brief description of our model, and the modelling results are given in Sect. \ref{sec:result}. 
A discussion and our conclusions are given in Sect. \ref{sec:discu}.

\section{\textit{XMM-Newton} observations and data reduction} \label{sec:data}

\subsection{XMM-Newton observations of PKS 2155-304}

\par PKS 2155–304 has been observed by the EPIC on board the \textit{XMM-Newton} satellite \citep{Struder2001A&A,Jansen2001A&A} on multiple occasions. For this work we focused on the EPIC-pn data, which have a larger effective area than EPIC-MOS. We searched the \textit{\XMM} Science Archive (XSA) for PKS 2155-304 and obtained a sample of 23 observations with EPIC-pn exposure. To study the intra-day variability, we further required the EPIC-pn exposure time to be $>$ 80 ks and obtained four observations (ObsIDs 0124930301, 0124930501, 0124930601, and 0411780101),
in which two observations (ObsIDs 0124930301 and 0124930601) each have two exposures, 
and the other two  observations (ObsID 0124930501 and 0411780101) have three exposures of similar duration in each observation with different filters. 
We dropped ObsID 0411780101 as it shows no significant variations, while significant variations with distinct patterns were   found in the other three observations. A flare (peak) is clearly visible in the LCs of ObsID 0124930601, while ObsIDs 0124930301 and 0124930501 show nearly monotonic decrease and increase, respectively, indicating the possible presence of strong flares that were not fully sampled in these two observations. 
Since the spectral and variability properties of the public archival data for individual exposures in these observations have been independently analysed by several authors \citep[e.g.][]{Gaur2010ApJ,Bhagwan2014MNRAS,Bhagwana2015FluxAS,Bhatta2025ApJ}, our aim was to obtain the lag-frequency spectra based on the combined LC in each observation.
The long densely sampled LCs with visible flares are particularly useful for performing high-quality timing analysis across a wide-temporal frequency range,
which could provide strong constraints on the jet properties and offer valuable insights into the underlying mechanisms at work in the blazar jet.
Meanwhile, we adopted an extra observation (ObsID 0411780701) as a reference or base state for modelling the SED; this observations has the lowest flux and shows no significant variation \citep{Bhagwan2014MNRAS, Gaur2017ApJ}.

\subsection{Data reduction}

\par Data reduction of these observations is a bit complicated as they all have multiple exposures performed in different modes or with different filters within a single observation. Specifically, ObsIDs 0124930301 and 0124930601 each have two exposures in Imaging and Timing modes, respectively, while ObsIDs 0124930501 and 0411780101 each have three exposures with the three different filters (Thin, Medium, and Thick). We carefully accounted for the complexity so that continuous and uniform light curves are available (see the method below).

\par The raw data were reduced with the \textit{XMM-Newton} Science Analysis System (SAS, version 20.0.0) and the Current Calibration Files. 
For data in Imaging mode, we filtered out time intervals with flaring background, and extracted the source light curves and spectra within a circular region with a radius of 60$\arcsec$, while extracting background from nearby source-free regions \citep[see][for details]{Kang_2024}. Nevertheless, significant pile-up effects were found in most observations by SAS task {\it epatplot}. We therefore employed an annular region with inner radius of 15$\arcsec$ for source products, which has been confirmed to be able to fully erase the pile-up effects for these observations. For the two exposures in Timing mode, we extracted the source products with RAWX in [27:47], while background in [3:5] \citep{Ng_2010, Hu2024ApJ}. There is no significant pile-up effect for the data in the Timing mode.

\par Furthermore, we reprocessed the light curves with a time bin of 100 s using the task {\it epiclccorr} in order to apply both relative and absolute corrections as well as the background subtraction. 
However, we note the transmittances of the filters would not be corrected by {\it epiclccorr},\footnote{\url{https://xmm-tools.cosmos.esa.int/external/sas/current/doc/epiclccorr.pdf}} 
which are similar for Thin and Medium, but quite different for Thick. Therefore, when comparing the light curves of the different filters in this work, we manually converted the observed count rates to intrinsic fluxes, where the conversion factors were chosen to make the mean flux of a light curve equal to that estimated with the corresponding spectrum. 

\par The spectra were re-binned to have at least 50 counts per bin with task {\it specgroup}, which were then fitted using XSPEC \citep{Arnaud_1996} with $\chi^{2}$ statistics in the 0.5--10 keV band. We find that a simple power law with Galactic absorption \citep[$N_{\rm H} = 2.5 \times 10^{20} \, \rm cm^{-2}$,][]{HI4PI_2016} could well fit the spectra ($\chi^{2}$/dof $\sim$ 1.0). With this simple model, we unfolded the spectra to derive the intrinsic energy distribution, after correcting the Galactic absorption. 
The broad-band X-ray spectra are plotted in Fig. \ref{figs:sed2} in Sect. \ref{sec:result}.

For each observation or exposure, we extracted LCs in two main energy sub-bands, 0.5–2 keV and 2–10 keV, which are the most frequently used in the spectral and temporal analysis of the X-ray variability. 
In addition, the full energy band was divided into four sub-bands, 0.5--1 keV, 1--2 keV, 2--4 keV, and 4--10 keV, in order to perform a detailed analysis of the inter-band variability characteristics.
All LCs (in units of $\rm erg\cdot cm^{-2} \cdot s^{-1}$) for ObsID 0124930301 are shown in Fig. \ref{figs:lc1},
while the LCs for the other two observations are presented in Figs. \ref{figs:lc2} and \ref{figs:lc3}.

\section{Fourier analysis}\label{sec:method}
Variability in blazars is often characterised by power spectral density (PSD) and time lags in the Fourier-frequency domain. 
In the following we describe how we performed the PSD and Fourier time-lag analyses of the variable LCs measured in three observations.

\subsection{Power spectral density}
The PSD provides a measure of variability power as a function of temporal frequency,
 and is a useful tool to search for the possible periodicities or QPOs in a LC. 
Particularly, \cite{Gaur2010ApJ} reported a weak QPO signal during the second \textit{\XMM} observation of PKS 2155-304 on 24 May 2002 (ObsID 0124930501),
based on an analysis of the X-ray PSD and structure function for the individual exposure data.

For the three observations analysed in this work, we estimated the PSDs by concatenating the LCs measured with different observation modes or filters, to 
 ensure that their PSDs could be closely aligned down to $\sim10^{-5}$ Hz.
Since there are gaps between the evenly sampled LCs measured using different observation modes or filters,
 we calculated the PSD using the Lomb-Scargle periodogram (LSP),
a widely used algorithm designed to detect and characterise periodic signals in unevenly sampled LCs \citep[see][and references therein]{VanderPlas2018ApJS}.
LSP is calculated using the \emph{lomb-scargle}\footnote{\url{https://docs.astropy.org/en/stable/timeseries/ lombscargle.html}} class provided by \emph{astropy}. 
The resultant PSDs for the concatenated LCs are shown in Fig. \ref{figs:psd}.

To characterise the type of noise present in the variations, we fitted the PSDs using the Bayesian approach described in \cite{Vaughan2010MNRAS}.
In this analysis, a power-law plus constant model is used to fit the PSDs of the variable X-ray LCs.
The best-fitting models are overplotted in Fig. \ref{figs:psd}.
In the figure we report the best-fitting power-law index $\alpha$
 and the frequency threshold $\nu_{\rm Pois}$ 
above which the variability is dominated by white noise resulting from the uncertainties in the measured count rate.
The results indicate that the intra-day PSDs of PKS 2155-304 can be well described by red noise, with an index varying around 2.0 below the frequency threshold $\nu_{\rm Pois}\sim(1-8.4)\times10^{-4}$ Hz.
These power-law indices of PSDs are not significantly different from those derived from the individual exposures reported in \cite{Gaur2010ApJ} and \cite{Bhatta2025ApJ}.

The variability of blazars can be described as a damped random walk (DRW) process \citep[e.g.][]{Zhang2022ApJ,Zhang2023ApJ},
a type of Gaussian process (GP) also known as the Ornstein-Uhlenbeck process in physics.
By employing the DRW model to simulate LCs, we can assess the significance level of any prominent peaks in the PSDs that 
 might suggest the presence of QPO behaviour in the time series data.
To this end, we simulated 5000 LCs for each sub-band using the \emph{celerite} algorithm,\footnote{\url{https://celerite.readthedocs.io/en/stable/}}
which offers a fast and flexible method for modelling LC with stochastic processes \citep{Foreman2017AJ}.
Based on the outcomes of DRW modelling, we derived the distribution of LSPs, and evaluated
 $1\sigma, 2\sigma$, and $3\sigma$ confidence contours. The results are presented in Fig. \ref{figs:psd}. 
A significant QPO signal can be claimed if a peak exceeds at least 3$\sigma$ above the red noise fit of the PSD.
Clearly, the PKS 2155-304 LSPs do not show any prominent peaks above 3$\sigma$ level in all the LCs considered in the work. 
Thus, our results indicate that no evidence for possible QPOs has been found in any of the PSD plots.

 \subsection{Frequency-dependent time lags} 
 
 The X-ray inter-band Fourier time lags and their associated errors were calculated following a standard approach reviewed in detail by \cite{Uttley2014AARv}.
 The procedure was implemented using the X-ray timing analysis package {\textit{pyLag}}.\footnote{\url{http://github.com/wilkinsdr/pylag}}

For each pair of LCs, we obtained the lag-frequency spectra in five logarithmically spaced bins, 
ranging from the lowest frequency $T^{-1}$ (where $T$ is the time duration of each observation) 
to the Nyquist frequency of $5\times10^{-3}$ Hz.
The resulting lag-frequency spectra are denoted as $\tau_1$.
To reveal the potential profile of the lag-frequency spectra, 
we then divided each LC into two segments of equal duration and recalculated the lag-frequency spectra, referred to as $\tau_2$. 
In these calculations the simple linear interpolation method was used to produce the missing data points in the concatenated LC for each energy band.
The uncertainties of the time lags in each frequency bin were given by the coherence, 
which is a measure of the fraction of the rms amplitude of one process at a given frequency that can be predicted from the other by a linear transform \citep{Vaughan1997ApJ}.

Alternatively, modelling the variability with GPs allows us to generate evenly sampled realisations of the LCs, 
from which the lags can be computed in the same frequency bins as in our previous analysis \citep{Wilkins2019MNRAS}.
As mentioned above, the DRW model was applied to fit the observed LCs, 
and the realisations were randomly drawn from the conditional posterior.
For each energy band, we drew 5000 evenly sampled realisations independently, and calculated the lag-frequency spectra for each pair of realisations.
The final lag-frequency spectra and 1$\sigma$ uncertainties were estimated as the mean and standard deviation of the resulting lag distribution in each frequency bin.

For the three observations of our interest, the lag-frequency spectra measured from the DRW model, and the linear interpolation are shown in Figs. \ref{figs:ftlag1}, \ref{figs:ftlag2}, and \ref{figs:ftlag3}, respectively.
The lags in all bands and frequency bins using DRW model are consistent with those computed from the linearly interpolated LCs.
In the figures, the green symbols represent the results obtained from the linear interpolation method,
 while the blue symbols represent the results obtained from the DRW model.

The measured lag–frequency spectra for ObsIDs 0124930301 and 0124930501 exhibit similar profiles, showing only the soft or negative lags;
 the magnitude of the lags increases as the frequency decreases.
Depending on the differences between the two compared energies, the magnitude of the lags at the lowest frequency is in the range  $\sim5-18$ minutes for ObsID 0124930301,
and  $\sim5-30$ minutes for ObsID 0124930501.
In ObsID 0124930601, the measured lags between different energy bands are characterised by a small hard or positive lag at the lowest frequencies, 
which drops to negative values at higher frequencies below $\nu_{\rm Pois}$. 
The magnitude of the soft lags is $\sim10-40$ minutes.
The measured soft and hard lags occur at frequencies well below the frequency threshold set by Poisson noise, 
and the magnitude of the lags between two compared LCs increases with the difference in energy.

Our results should be consistent with those reported in \cite{Zhang2006ApJ}, where the X-ray inter-band time lags were measured using the popular cross-correlation function (CCF) method \citep{Peterson1998PASP}.
On the basis of individual exposures, \cite{Zhang2006ApJ} estimated the time lags with different techniques for two cases: 0.2--0.8 keV versus 0.8--2.4 keV (i.e. soft vs. medium energy band) and 0.2–0.8 keV versus 2.4–10 keV (i.e. soft vs. hard energy band).
Using LCs measured in Timing mode in ObsID 012940301, the authors found that $\tau_{\rm cent}$ (representing the CCF centroid) is $\sim -2.5$ minutes for both cases,
and $\tau_{\rm fit}$ (representing the peak of an asymmetric Gaussian function fitting the CCF) is $\sim-7.5\pm2.5$ and $-15\pm4$ minutes for the soft--medium and the soft--hard CCFs, respectively.
For exposure with Thin filter in ObsID 012940501,  
$\tau_{\rm cent} \sim -27\pm5$ and $\sim-47\pm7$ minutes, and $\tau_{\rm fit }\sim -14\pm 3$ and $\sim-23\pm6$ minutes for the soft--medium and the soft--hard bands, respectively.
For exposure in Imaging mode in ObsID 012940601, soft lags of about 0.5 and 1 hr were derived for the soft--medium and soft--hard CCFs, respectively.

The detected soft lags may be consistent with the very steep X-ray spectrum observed in the three epochs.
Such steep spectra are indicative of the X-ray emission originating farther above the synchrotron peak frequency,
where the electron populations are dominated by the synchrotron cooling process.

\section{Model} \label{sec:model}
To extract the physical information contained in the X-ray emission,
it is essential to perform a detailed theoretical modelling of the time-dependent electron acceleration and radiation transport processes in the jets.
Below, we provide a brief overview of the model proposed in \cite{Hu2024ApJ}.

In the model, the observed X-rays and TeV $\gamma$-rays are assumed to be predominantly composed of two emission components: 
a non-stationary (active) component and a quasi-stationary component. 
The former is responsible for the short-timescale IDV,  while the latter changes only on longer timescales.
For simplicity, the two components are decoupled from each other, and are characterised by different physical conditions.
Each component is defined by ten physical quantities (i.e. $B'$, $\delta_{\rm D}$, $R'$, $a_{\rm sh}$, $\nu_{\rm pk}$, $\eta'_{\rm esc}$, $q_0'$, $\gamma_{\rm i,min}'$,  $\gamma_{\rm i,max}'$, and $p$).
Hereafter, quantities measured in the co-moving frame are denoted with primes, while quantities in the stationary AGN frame are unprimed, unless specified otherwise.
The first three quantities describe the global properties of the zone, including the magnetic field strength, Doppler factor, and spatial size.
The quantities $a_{\rm sh}$, $\nu_{\rm pk}$, and $\eta'_{\rm esc}$ determine the ratio of shock to stochastic acceleration, the observed synchrotron peak frequency and electron escape efficiency, respectively. 
For each zone, the injected electrons follow a power-law distribution characterised by the last four quantities:  the injection rate,
the minimum and maximum Lorentz factors, and the power-law index, represented by
$q'_{\rm e}(\gamma')=q_{\rm 0}'\gamma'^{-p} H(\gamma';\gamma'_{\rm i, min},\gamma_{\rm i, max}')$, 
where $H(\cdot)$ is the Heaviside function defined by $H=1$ if $\gamma'_{\rm i, min}\le\gamma'\le\gamma_{\rm i, max}'$, and $H = 0$ otherwise.

After injection, the evolution of the emitting electrons in both zones can be independently described by the transport equation \citep[e.g.][]{Schlickeiser1984,Stawarz2008,Hu2023ApJ}
\begin{eqnarray}\label{acc-transport1}
\frac{\partial N'_{\rm e}(\gamma',t')}{\partial t'} &=& \frac{\partial}{\partial \gamma'} \left\{\gamma'^2D_p(\gamma')  \frac{\partial}{\partial \gamma'}\left[\frac{N'_{\rm e}(\gamma',t')}{\gamma'^2}\right]  \right\} \nonumber\\
&-& \frac{\partial}{\partial \gamma'}\left[\left(A_0^{\rm sh}\gamma'+\dot\gamma'_{\rm rad}\right) N'_{\rm e}(\gamma',t') \right] \nonumber\\
&-& \frac{N'_{\rm e} (\gamma',t')}{t'_{\rm esc}} + Q'_{\rm e}(\gamma',t').
\end{eqnarray}

\begin{table*}
        \centering
        \caption{Input and derived quantities from the modelling the observed X-ray spectra and Fourier time lags with our proposed two-zone model.}
        \label{tabs:twozones}
        \begin{tabular}{cccccccccccc} 
                \hline\hline
                ObsID            &$\delta_{\rm D}$      &       $B'$      & $R'$         & $\nu_{\rm pk} (E_{\rm pk})$   & $q_0'$     & $\gamma'_{\rm i,max}$  & $p$ &   $\eta'_{\rm esc}$& $t'_{\rm acc}$   & $D_0$  & $U_{\rm e}'/U_{\rm B}'$  \\
                                 &                              &       G           &    $10^{15}$cm             &$10^{17}$ Hz   (keV)                   & $\rm s^{-1}$             &$10^5$        &  &     &$R'/c$   &$10^{-7}\rm s^{-1}$      \\
                \hline
        0411780701      &16             &0.06            &78    &$0.1(0.06)$ &$1.26\times10^{48}$                    &$10$   &$2.38$ & $1.00$&$1.48$ &$0.65$ & 3.2             \\ 
        0124930301      &50             &0.12            &3.0           &$3.0 (1.24)$ &$1.89\times10^{46}$            &$2.0$  &$2.3$  & $1.40$&$4.39$ &$5.69$ & 35.3     \\ 
        0124930501      &45             &0.11    &2.0           &$4.5 (1.86)$ &$2.10\times10^{46}$            &$2.8$  &$2.3$  & $1.40$&$5.81$ &$6.44$ & 149.0    \\ 
        0124930601      &40             &0.2             &1.0           &$2.3 (0.95)$ &$3.82\times10^{46}$            &$2.1$  &$2.45$ & $1.26$&$6.26$ &$11.98$        & 95.9    \\ 
                \hline\hline
        \end{tabular}
\end{table*}

Considering  hard sphere scattering for interactions with MHD waves, the momentum diffusion coefficient $D_p (\gamma')$ is given by
\begin{equation}
D_p(\gamma') = D_0\gamma'^2.
\end{equation}
This makes the acceleration timescale independent of energy. Then, the normalisation of the diffusion coefficient  can be evaluated by the relationship
\begin{equation}
D_0=\frac{1}{(4+a_{\rm sh})t_{\rm acc}'},
\end{equation}
where $a_{\rm sh}\equiv A_0^{\rm sh}/D_0$ with $A_0^{\rm sh}$ accounting for first-order Fermi acceleration, and the energy-independent acceleration timescale is 
\begin{equation}\label{eqs:tacc}
t'_{\rm acc}=\frac{6\pi m_{\rm e}c}{\sigma_{T}B'^{3/2}}\sqrt{\frac{\nu_0 \delta_{\rm D}}{\nu_{\rm pk}(1+z)} },
\end{equation}
which is determined by the balance of the synchrotron cooling and acceleration.
Here, $\nu_0=4m_{\rm e}c^2/3hB_{\rm cr}(1+z)$, with $m_{\rm e}$, $c$, $h$, and $B_{\rm cr}$ being the rest mass of electrons, 
light speed, Planck's constant, and critical magnetic field strength, respectively.

In the model the spatial transport of the electrons in the blob is parametrised  through an energy-independent escape timescale $t'_{\rm esc}=t'_{\rm acc}/\eta'_{\rm esc}$. 
We consider synchrotron cooling to be the dominant energy losses governing the evolution of the electrons emitting X-rays,
and therefore the term $\dot\gamma'_{\rm rad}=-(\sigma_{\rm T}B'^2/6\pi m_ec)\gamma'^2$ (where $\sigma_{\rm T}$ is the Thomson cross-section).

Since the X-ray and TeV $\gamma$-ray fluxes in HBLs are strongly correlated, and very highly variable when the electron injection rate changes.
We assume that the last term $Q'_{\rm e}(\gamma',t')$ is a time-dependent function that describes a time-dependent injection process of the energetic electrons. 
The term can be decomposed into the product of $f_{\rm e}'(t')$ and $q'_{\rm e}(\gamma')$, 
with $f'_{\rm e}(t')$ and $q'_{\rm e}(\gamma')$ being the time-dependent profile and time-independent energy distribution, respectively.

Given all the parameters above, the time-dependent electron energy distribution (EED) can be obtained by solving the diffusion equation (Eq. \ref{acc-transport1}) numerically with the implicit Crank-Nicolson scheme, 
and then we calculate the synchrotron and SSC radiation using standard formulas in the literature\citep[e.g.][]{Jones1968,Deng2021MNRAS}. 
The theoretical LCs and Fourier time lags can be calculated numerically following the methods described in \cite{Hu2024ApJ} and references therein.


     \begin{figure}
   \centering
     \includegraphics[width=0.45\textwidth]{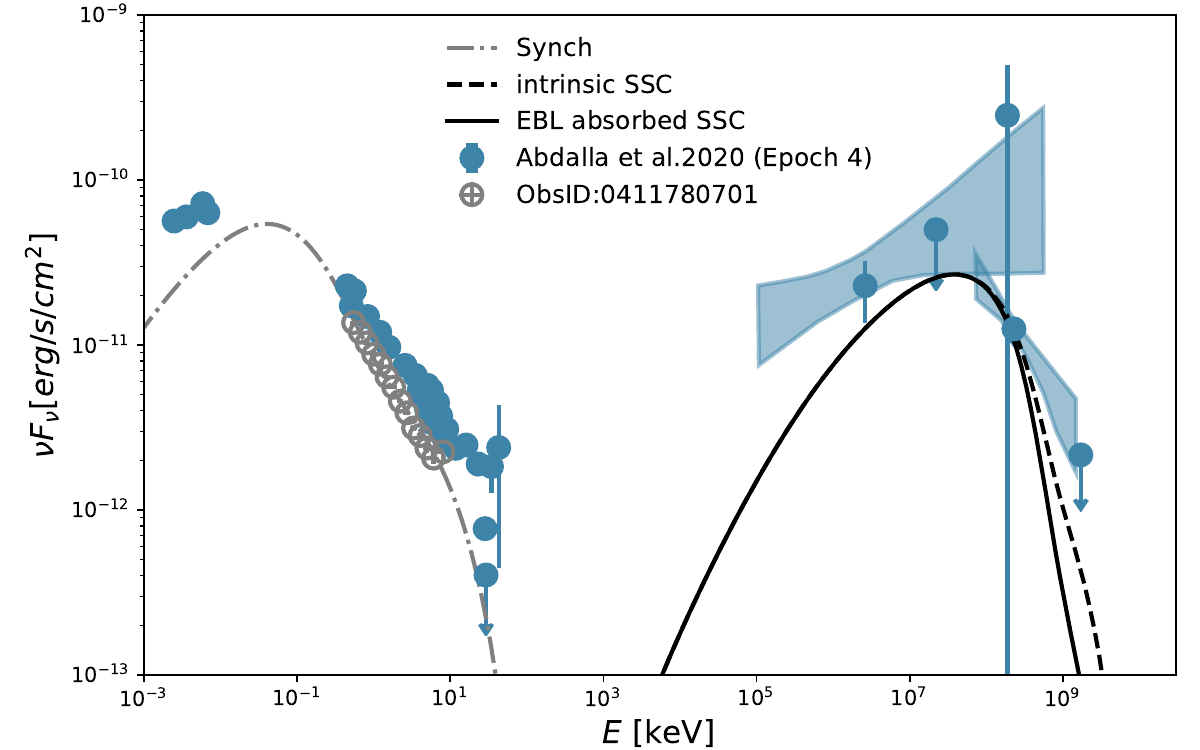}
   \caption{Modelling the lowest X-ray flux from the quasi-stationary zone. 
   }
              \label{figs:bgsed}%
    \end{figure}

 \begin{figure}
   \centering
  \includegraphics[width=0.45\textwidth]{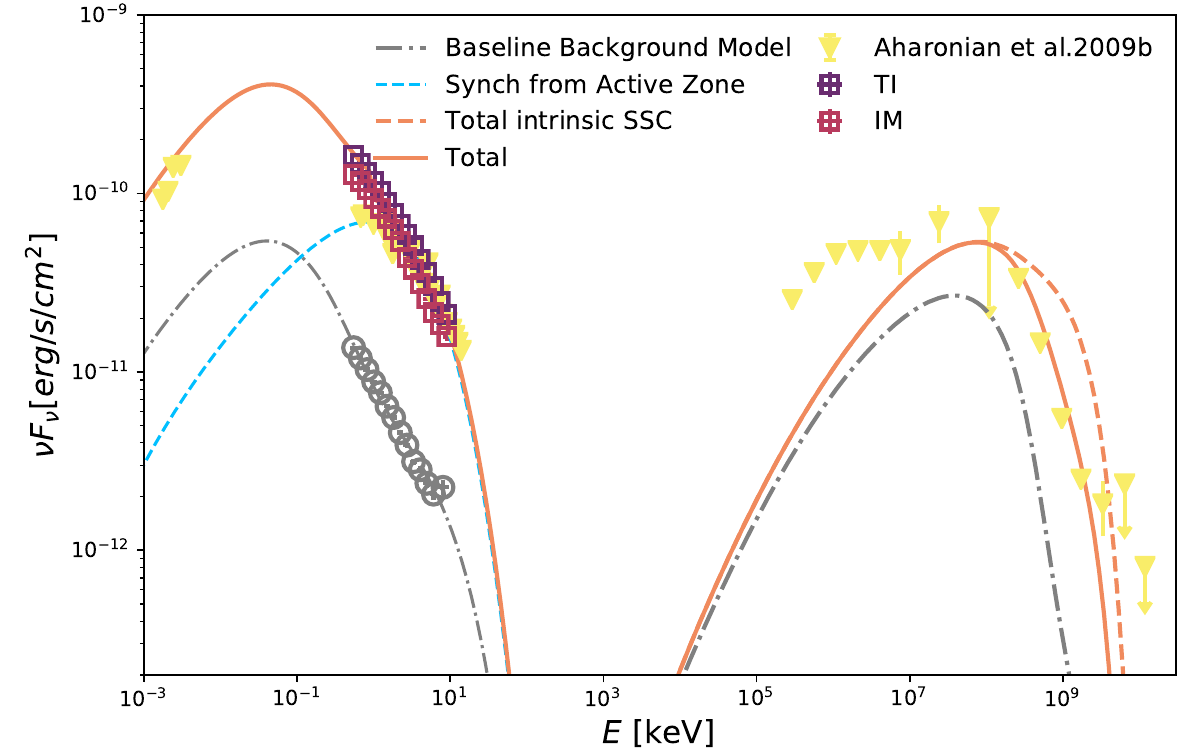}
  \includegraphics[width=0.45\textwidth]{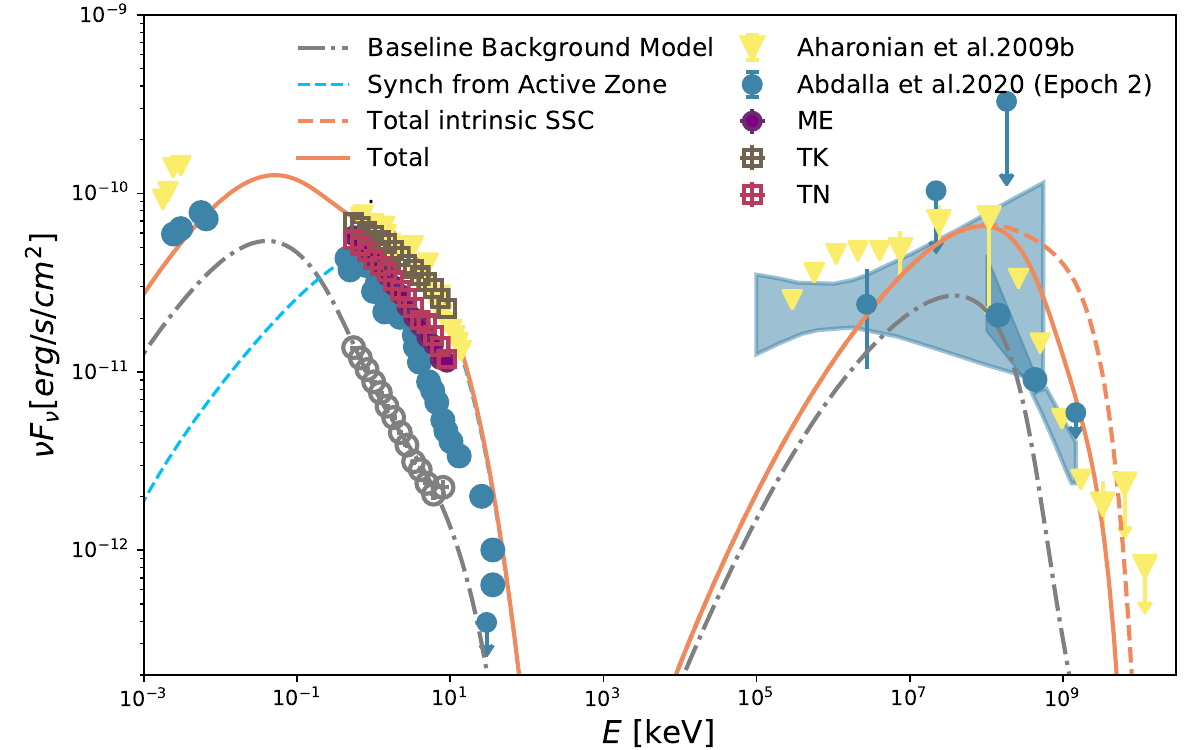}
  \includegraphics[width=0.45\textwidth]{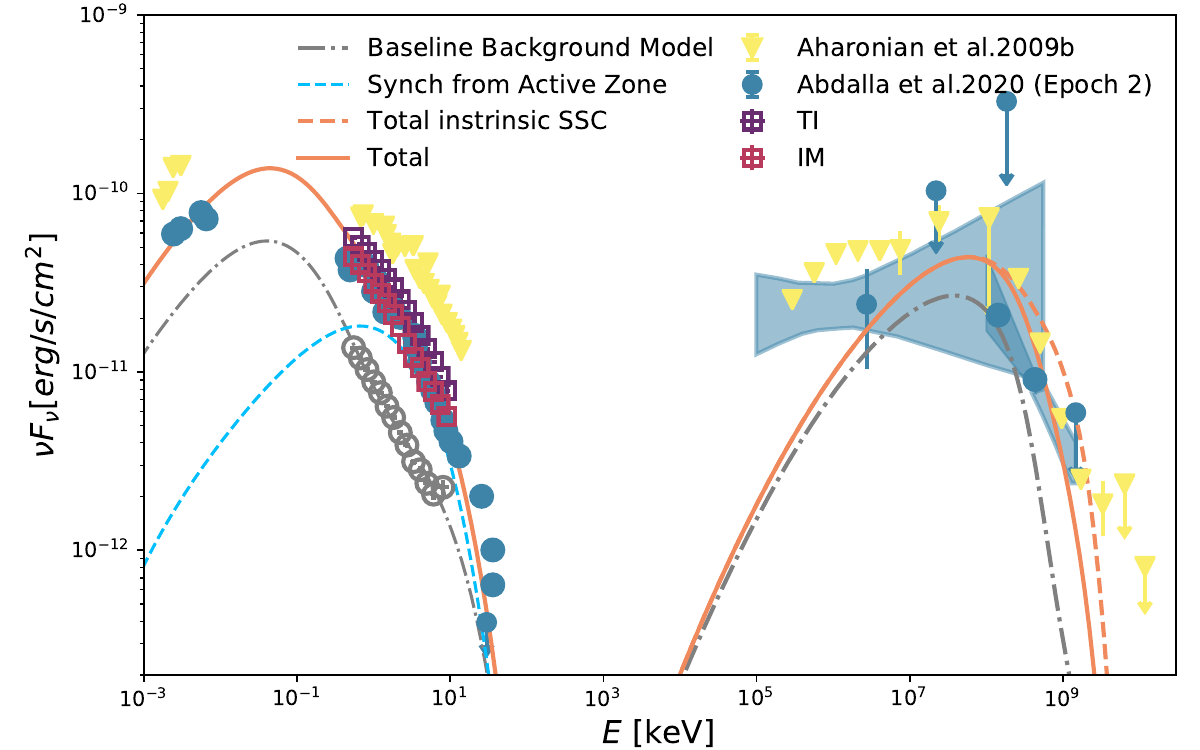}
   \caption{Modelling the observed X-ray spectra (Top: 0124930301; Middle: 0124930501; Bottom: 0124930601) with our proposed two-zone leptonic model. 
      The various components are labelled in the legends.
      Additionally, we mark the contemporaneous data from \cite{Aharonian2009ApJ} and \cite{Abdalla2020A&A}, which serve as valuable references for our modelling. 
      The grey empty circles are the lowest X-ray flux  observed in ObsID 0411780701.  
        }
    \label{figs:sed2}%
    \end{figure}

    \begin{figure*}
   \centering
  \includegraphics[width=0.8\textwidth]{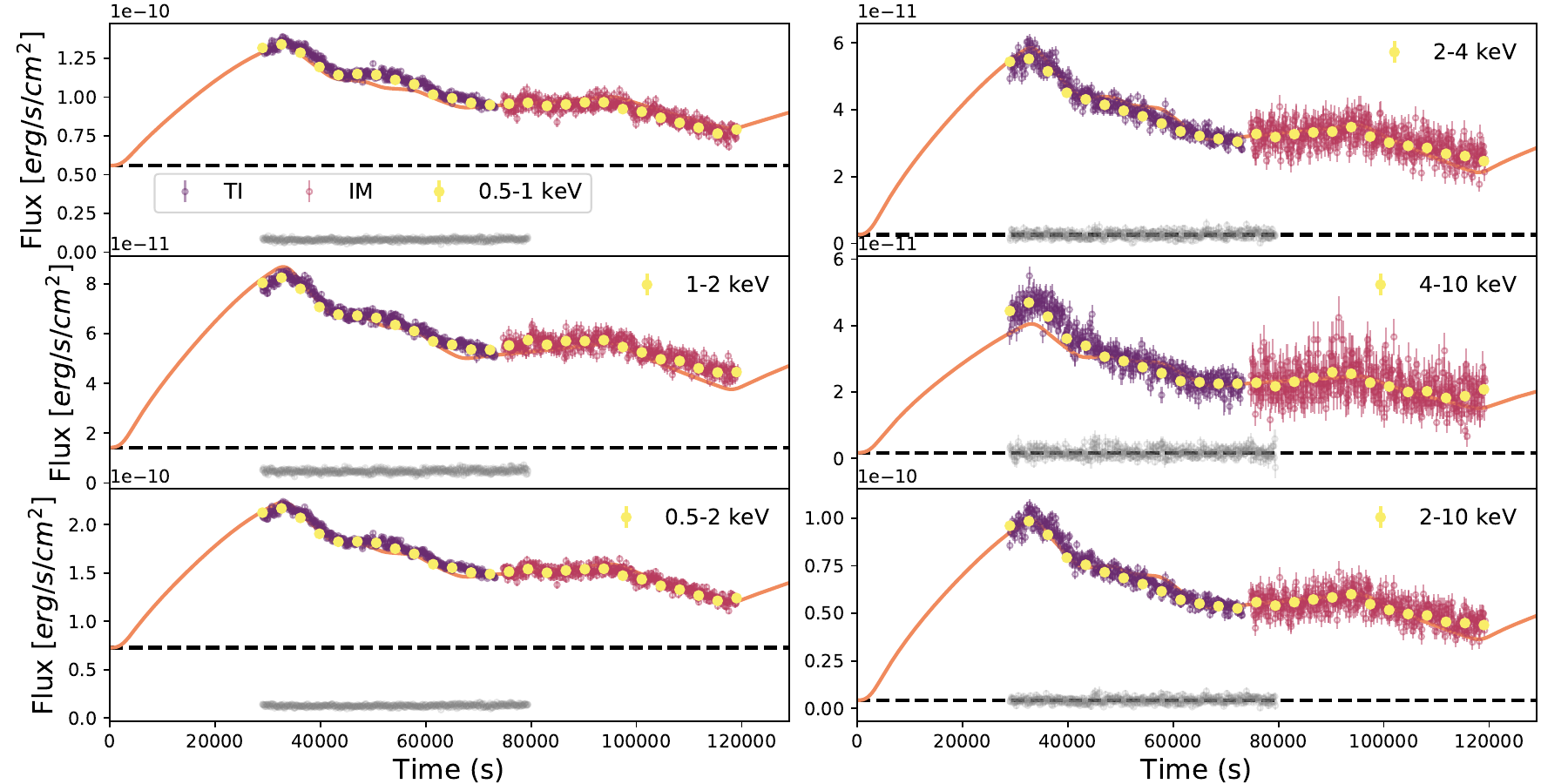}
   \caption{\textit{XMM-Newton} EPIC-pn LCs from ObsID 0124930301. TI and IM represent the data in Timing and Imaging science modes, respectively.  The yellow data points denote the 1 hr binned LCs.
    The grey circles denote the lowest X-ray flux observed in ObsID 0411780701.  
   The dashed lines represent the flux level from the quasi-stationary zone. The orange solid lines represent the theoretical LCs re-produced by our two-zone model.}
              \label{figs:lc1}%
    \end{figure*}

\section{Results}\label{sec:result}

In an SSC framework, the electrons responsible for the X-ray emission also produce the TeV $\gamma$-ray emission.
Due to the lack of simultaneous TeV $\gamma$-ray data corresponding to our selected observations, 
the simultaneous MWL SEDs constructed in 2008 by \cite{Aharonian2009ApJ} and in 2013 by \cite{Abdalla2020A&A} are employed as a reference.
With respect to the campaign carried out in 2008, the source was found in a low flux state during the 2013 campaign, and exhibited highly significant flux variability in the X-rays.
We note that the X-ray fluxes of PKS 2155-304 observed in ObsID 0124930301 are compared with those observed during the 2008 campaign,
while the X-ray fluxes measured in ObsID 0124930601 are compared with those observed in Epoch 2 in the 2013 campaign.
For ObsID 0124930501, the X-ray flux is between those observed in Epoch 2 and the 2008 campaign.
In addition, we find that the lowest X-ray flux measured during ObsID 0411780701 is slightly lower than that observed in Epoch 4 in the 2013 campaign. 
Thus, the contemporaneous TeV $\gamma$-ray data collected during these two campaigns are considered to estimate the TeV $\gamma$-ray flux produced in the observations in this work.
We note  that the intrinsic TeV $\gamma$-ray emission is expected to be attenuated by photon-photon interactions with the extragalactic background light (EBL) photons in the optical-to-IR waveband. 
The EBL model by \cite{Franceschini2008A&A} is adopted to correct for EBL absorption.

In addition, the observed UV data during these two campaigns are also considered in order to constrain the model parameters.
The UV radiation should be emitted from larger-scale regions in the jet \citep{Blandford1979,Konigl1981}.
Using 20 \textit{XMM-Newton} observations of PKS 2155-304 during the period of 2000–2012,
 \cite{Bhagwan2014MNRAS} found that the optical/UV variations are not correlated with those in the X-ray, 
 and the log-parabolic model with an additional power-law component is required to model the simultaneous optical/UV--X-ray SED.
These suggest that the optical/UV and X-ray emissions may arise from different parts of the jet. 
In our modelling, we require that the total emission from the quasi-stationary and active zones cannot exceed the observed UV data.

Because the X-ray emission (0.5–10 keV) of PKS 2155-304 lies mostly in the falling part of the low-energy bump,
which is commonly believed to be the synchrotron from the most energetic electrons spiralling around the magnetic fields in relativistic jets,
the values of $\gamma_{\rm i,min}$ cannot be effectively constrained.
In the course of the fitting, we fixed $\gamma'_{\rm i,min} = 2\times10^3$, which is close to the particular characteristic energy scale $\sim m_{\rm p} /m_{\rm e}$ introduced by a
proton-mediated shock scenario \citep{Ushio2010,Abdo2011ApJ501}.
Meanwhile, $a_{\rm sh}=0$ is also held fixed. 
This is motivated by the fact that the X-ray spectra of PKS 2155–304 show curvature or deviation from a single power law and can be well modelled by a log parabola function, 
and the curvature of the X-ray spectra is anti-correlated with the peak energy \citep{Massaro2008A&A,Gaur2017ApJ,Bhatta2025ApJ}.
These may be the signals of stochastic acceleration.
Moreover, we start our modelling by setting $\delta_{\rm D}=\Gamma=50$, to produce the rapid variability and overcome the photon opacity barrier.
An extremely high bulk Lorentz factor $\Gamma>50$ of the jet may be difficult to be reconciled with the currently favoured models of jet acceleration \citep{Begelman2008MNRAS,McKinney2009MNRAS}.
Hence, we only consider $\delta_{\rm D}\lesssim50$ as a reasonable solution in our modelling.

Following the modelling strategy adopted in our previous work, we first need to estimate the quasi-stationary emission that remains constant during the fast variability.
For this purpose, the baseline emission of the quasi-stationary zone is estimated by modelling the lowest X-ray spectrum measured in ObsID 0411780701, 
while the collected simultaneous UV, X-ray, and TeV $\gamma$-ray  data from Epoch 4 of the 2013 campaign are used to guide our modelling.   
Assuming that there is no significant variation in the spectral shape of the emission from the quasi-stationary zone in different epochs,
we can use the theoretical SED to decompose the observed X-ray spectrum measured in ObsIDs 0124930301, 0124930501, and 0124930601. 
In the modelling, we fit the observed X-ray spectrum from these three observations, 
using the steady-state solution generated by continuously injected electrons.\footnote{Here we set the time-dependent profile of injection $f'_{\rm e}(t')=1$,
and then the steady-state solution ${\partial N'_{\rm e}(\gamma',t')}/{\partial t'}=0$ is obtained using a bidirectional Runge-Kutta method developed by \cite{Lewis2018ApJ}. } 
On the other hand, 
the theoretical LCs in different energy bands are generated by applying a constant injection with a duration of $2R'/c$, represented by a step function: $f'_{\rm e}(t')=1$ for $t' \in[0, 2R'/c]$,  otherwise $f'_{\rm e}(t')=0$. This is because the phase lags measured from the argument of the cross-spectrum between two energy channels are independent of the injection profile, when variations in flux are caused by the injection of energetic electrons.
The associated lag-frequency spectra can then be calculated numerically using the discrete Fourier transform \citep{Hu2024ApJ}.

In Fig. \ref{figs:bgsed} we display the theoretical SED corresponding to the lowest flux, together with the observed data. 
The theoretical SEDs and observed data in ObsIDs 0124930301, 0124930501, and 0124930601 are shown in Fig. \ref{figs:sed2},
and the theoretical lag-frequency spectra between various sub-bands are plotted in Figs. \ref{figs:ftlag1}, \ref{figs:ftlag2}, and \ref{figs:ftlag3}, as denoted by the orange lines.
All model parameters are reported in Table \ref{tabs:twozones}.
It can be seen that our model satisfactorily describes all of the measured lag-frequency spectra, 
and the theoretical X-ray spectra agree fairly well with the observations.

Since PKS 2155-304 in ObsID 0411780701 was in a rather low state with an average flux of $\sim4.3\times10^{-12} \rm ~erg \cdot cm^{-2} \cdot s^{-1}$ in 2--10 keV band,
it is reasonable to assume that the jet plasma approaches equipartition based on the minimum energy argument.
In modelling the X-rays in ObsID 0411780701, the model parameters are adjusted to ensure that the magnetic field and electron energy densities tend toward equipartition, 
while considering only the parameter space that satisfies the conditions $t'_{\rm esc}>t'_{\rm dyn}$ and the Alfv\'{e}n velocity $\beta_{\rm A}\lesssim1$ \citep{Hu2021MNRAS,Hu2023ApJ}.

Once the theoretical background emission from the quasi-stationary zone was determined, we initially carried out a rough reproduction of the SEDs of the active zones.
Subsequently, the Fourier frequency-dependent lag spectra were reproduced by adjusting the parameters that are related to synchrotron loss ($B'$), stochastic acceleration ($D_0$ or $\nu_{\rm pk}$), 
and the distribution of injected electrons ($\gamma'_{\rm max}$ and $p$). Following this, we readjusted the fitted parameters to better reproduce the SEDs. 
By iteratively repeating these procedures, we eventually obtained the parameters of the active zones in PKS 2155-304 for three different epochs, as shown in Table \ref{tabs:twozones}.

For the time-dependent model considered here, a minimum $\chi^2$ approach for estimating the optimum values of the model parameters and their statistical significance is potentially unfeasible because of expensive computational costs and the substantial number of adjustable parameters involved. 
In future work, we will implement a rigorous $\chi^2$ minimisation strategy to precisely determine the uncertainties on the fit parameters by improving the algorithms and models.

In the fitting procedure of the lag-frequency spectra, we mainly focus on the lags measured between the hard (2–10 keV) and soft (0.5–2 keV) bands.
We find that all other lag-frequency spectra measured from the observed LCs can be naturally explained.
Our results predicted that positive lags occur at the lowest frequency of the theoretical lag-frequency spectra in ObsID 0124930301 and 0124930501.
These are similar to the results for ObsID  0124930601. 
The absence of positive lags in the observed results may be due to the possible presence of strong flares not being fully sampled in these two observations.
Meanwhile, to interpret the corresponding broad-band X-ray SED, we find that the 0.5--2 keV background emission represents about 45\%, 42\%, and 42\% of the observed mean fluxes for ObsIDs 0124930301, 0124930501, and 0124930601, respectively, assuming that the X-rays above 2 keV are the sum of emission from the active zone and the baseline flux from the quasi-stationary zone.

After determining the parameters of the active zones, we  reproduced the observed LCs at two main sub-bands. 
Since the precise form of injection, which may reflect the underlying physics driving the variability, is not well known,
for simplicity, a series of step functions was employed to model the injection process.
We initiated our modelling by introducing a single step function, and then improved the fitting results by gradually increasing the number of step functions.
Finally, the LCs of two main sub-bands observed in three different epochs were satisfactorily reproduced by taking into account 8, 8, and 13 step functions.
The duration of a single injection ranges from $t'_{\rm dyn}$ to $13t'_{\rm dyn}$. 
In Fig. \ref{figs:lc1} we display the comparison between the theoretical and observed LCs for the first observation.
The comparisons for the other two cases are provided in Figs. \ref{figs:lc2} and \ref{figs:lc3}.
The results demonstrate that it is reasonable to assume that the background radiation remains approximately constant on an intra-day timescale.
We note that it is difficult to predict the temporal variability characterised by  random aperiodic flux fluctuations.
However, the variability in the Fourier frequency domain may be consistent with the red-noise-type processes potentially linked to the electron injection.

\begin{figure}
   \centering
  \includegraphics[width=0.45\textwidth]{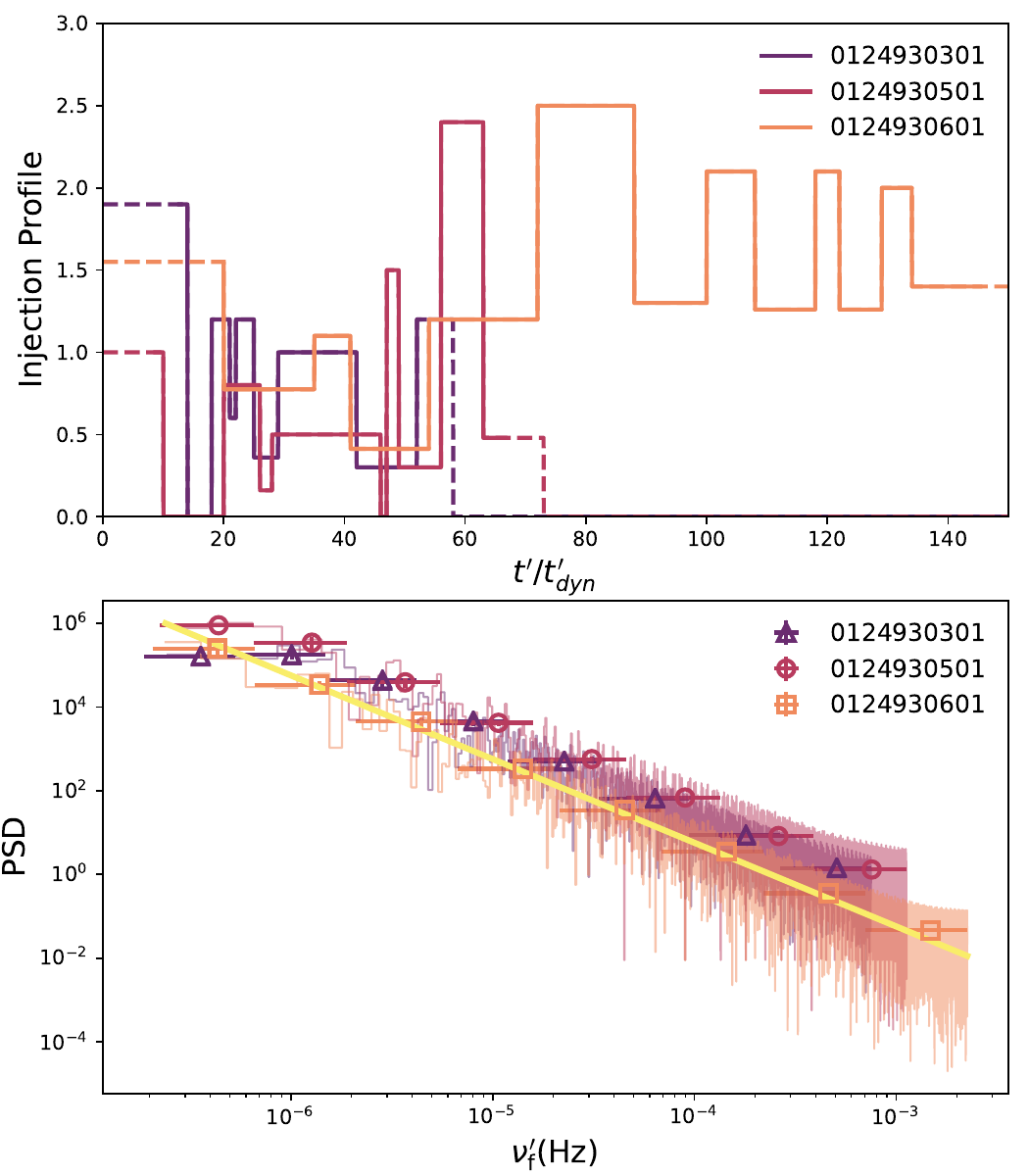}
   \caption{Upper panel: Time-dependent profile of the injected electrons  derived from the modelling of the LCs. 
   The portions represented by the solid lines correspond to the injections associated with the observed data.
Lower panel: Corresponding PSDs.  The step-shaped lines illustrate the `raw' periodogram. The symbols denote the logarithmically binned versions of the PSDs.
The thick yellow line shows the pure power-law PSD with an index of $-2$ (i.e. $P(\nu'_{\rm f})\propto {\nu'_{f}}^{-2}$). }
              \label{figs:profiles}%
    \end{figure}

For illustration, the injection profiles are given in the upper panel of Fig. \ref{figs:profiles}.
These results reveal that the underlying injection process may be very complex.
Furthermore, our findings indicate that the PSDs of the injections for the three observations are in agreement with the red-noise-type correlated stochastic processes. The results, which are consistent with the PSD analysis of flux variations, are presented in the lower panel of Fig. \ref{figs:profiles}.

By modelling the observed LCs in the 0.5--1 keV and 1--2 keV bands, we further find that the 0.5--1 keV and 1--2 keV background emissions
 are about 55\% and 24\% of the mean fluxes observed in ObsID 0124930301.
For ObsID 0124930501, about 52\% and 33\% of the mean fluxes in the 0.5--1 keV and 1--2 keV bands are attributed to the quasi-stationary emission zone,  
whereas the 0.5--1 keV and 1--2 keV background emissions are respectively about 53\% and 30\% of the mean fluxes observed in ObsID 0124930601.
The results are consistent with those obtained in 0.5-2 keV band.
The inferred quasi-stationary background emission in the 0.5--1 and 1--2 keV bands should be reasonable because lower energy flux should be emitted by a larger region than which produces higher energy flux.
It should be pointed out that the SEDs observed in three different epochs in Fig. \ref{figs:sed2} are reproduced using the refined background emissions,
which provide a more accurate approximation of the 0.5--1 and 1--2 keV X-rays SEDs in comparison with those inferred from the whole soft X-ray band.

Motivated by the successful modelling of the broad-band SED, lag-frequency spectra, and LCs,
we suggest that the X-ray IDV can be attributed to the injection of energetic electrons, while the other physical properties of the jet could remain constant on IDV timescales.
The energetic electrons may be associated with highly efficient particle acceleration, either due to the formation of shocks and turbulence in the jet flow, 
or annihilation of magnetic field lines at magnetic reconnection sites \citep[e.g.][]{Cerutti2012ApJ,Comisso2018PhRvL,Petropoulou2019ApJ,Marcowith2020LRCA}.

In our model, the injected electrons are then further accelerated by stochastic interactions with turbulent fluctuations, which is critical to determine the inter-band lag-frequency spectra.
Our modelling results show that the inferred diffusion coefficient $D_0$ ranges from $\sim10^{-7}$ to $\sim10^{-6}~\rm s^{-1}$,
which is comparable with the results inferred from other HBLs \citep[e.g.][]{Lewis2016,Hu2021MNRAS,Hu2023ApJ} and Fermi bubbles \citep{Sasaki2015ApJ}.

\begin{table*}
        \centering
        \caption{Input and derived physical quantities from the modelling of the observed X-ray spectra and Fourier time lags with a one-zone SSC model.}
        \label{tab:sed1}
        \begin{tabular}{cccccccccccc} 
                \hline\hline
        ObsID    &$\delta_{\rm D}$      &       $B'$      & $R'$        & $\nu_{\rm pk} (E_{\rm pk})$     & $q_0'$     & $\gamma'_{\rm i,max}$  & $p$ &   $\eta'_{\rm esc}$& $t'_{\rm acc}$   & $D_0$  & $U_{\rm e}'/U_{\rm B}'$ \\
                                 &                              &       G           &    $10^{15}$cm             &$10^{17}$ Hz   (keV)                   & $\rm s^{-1}$             &$10^5$        &  &     &$R'/c$   &$10^{-7}\rm s^{-1}$     \\
\hline
        0124930301      &50             &0.5             &5.5           &$0.06 (0.025)$ &$5.16\times10^{47}$   &$1.7$  &$2.50$ & $1.47$&$1.99$ &$6.84$ & 0.58    \\ 
        0124930501      &50             &0.3             &6.5           &$0.02 (0.008)$ &$1.54\times10^{47}$   &$4.5$  &$2.45$ & $1.40$&$6.3$  &$1.84$ & 1.59    \\ 
        0124930601      &50             &0.2             &8.2           &$0.05 (0.021)$ &$2.31\times10^{47}$   &$3.0$  &$2.50$ & $1.47$&$5.78$ &$1.58$ & 2.42    \\  
\hline\hline
        0124930301      &50             &0.03            &63    &$0.15 (0.06)$ &$3.34\times10^{47}$    &$9.0$  &$2.45$ & $1.47$&$7.48$ &$0.16$ & 8.04  \\ 
        0124930501      &50             &0.03            &50    &$0.03(0.01)$ &$2.20\times10^{46}$    &$12.0$         &$2.25$ & $1.40$&$21.08$        &$0.07$ & 12.30   \\ 
        0124930601      &50             &0.07            &23    &$0.07(0.029)$ &$1.59\times10^{47}$    &$5.0$  &$2.45$ & $1.47$&$8.42$ &$0.39$ & 4.82  \\ 
\hline\hline
        \end{tabular}
        \tablefoot{The upper and lower panel show Case A and Case B, respectively.}
\end{table*}    
    
\section{Discussion and conclusions}\label{sec:discu}
The modelling of broad-band SEDs and the variability of blazars is very important in understanding the jet physics and discriminating among the different models.

\begin{figure}
   \centering
  \includegraphics[width=0.45\textwidth]{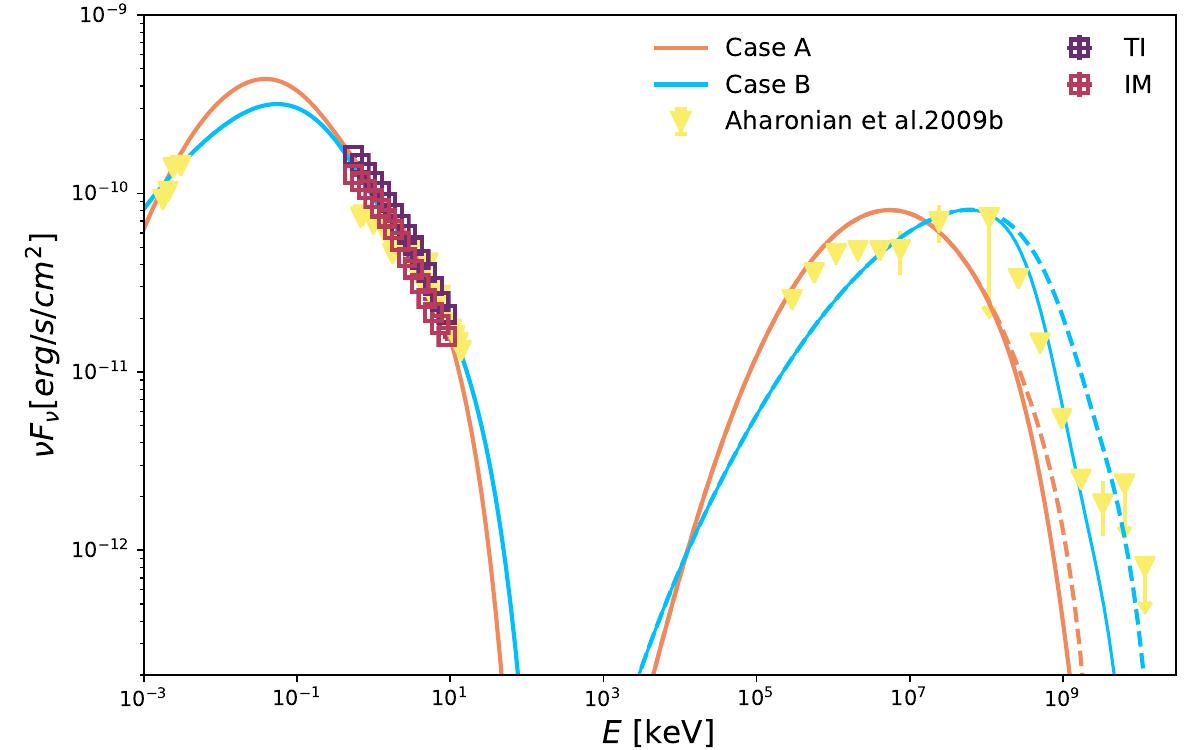}
    \includegraphics[width=0.45\textwidth]{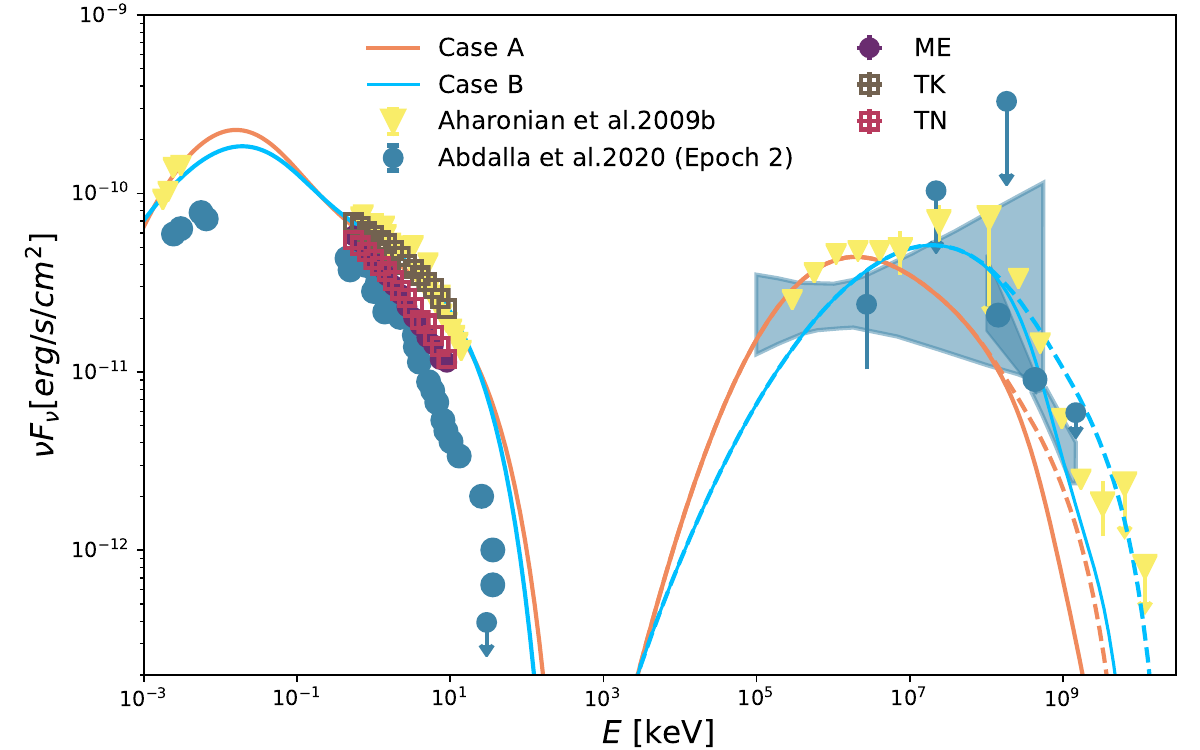}
    \includegraphics[width=0.45\textwidth]{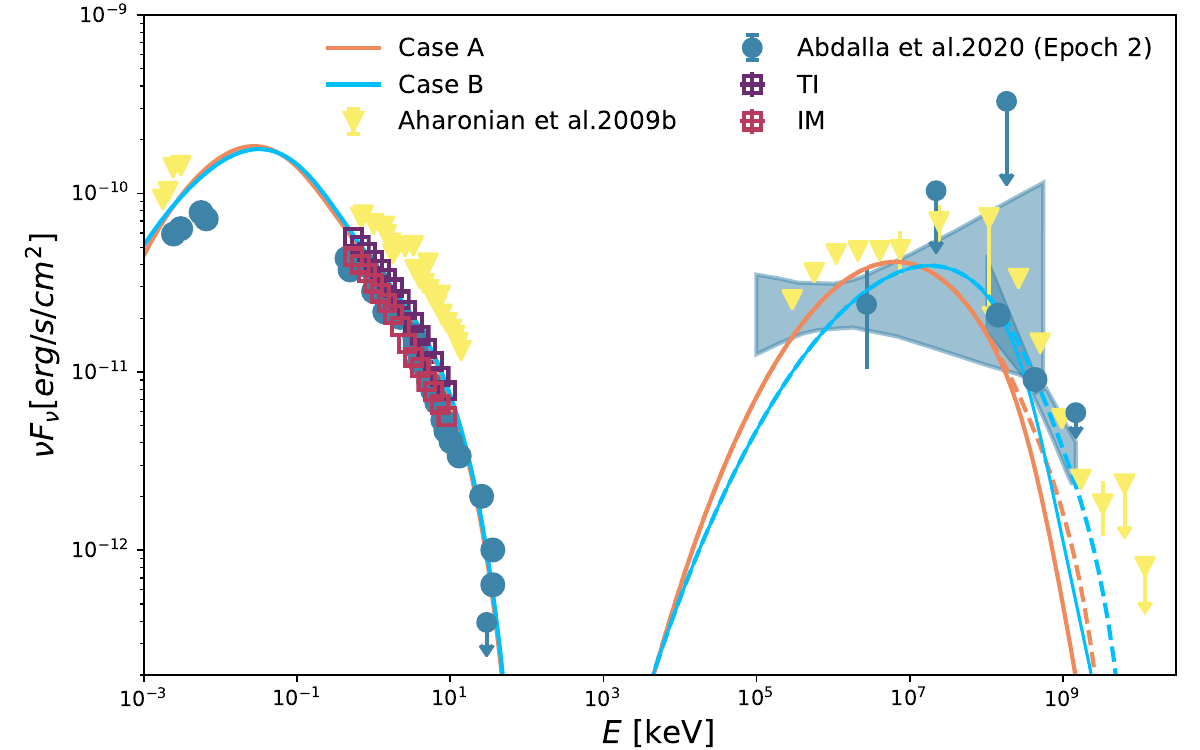}
   \caption{Modelling the observed SEDs with the classical one-zone SSC model.
    The dashed lines are the intrinsic SSC radiation.
    }
    \label{figs:sed1}%
    \end{figure}

The simultaneous MWL SEDs of HBLs are generally studied using classical one-zone homogeneous SSC models \citep[e.g.][]{Ghisellini2014Natur,Chen2018ApJS,Deng2021MNRAS},
which have successfully reproduced many blazar SEDs and provided definite predictions for the flux and spectral variability that should be seen in the low- and high-energy peaks.
In the context of a single-zone SSC model, the overall UV, X-ray, and $\gamma$-rays originate from a single compact region.
To illustrate this, the resultant SEDs are presented in Fig. \ref{figs:sed1}, with the parameters reported in Table \ref{tab:sed1}.
In the absence of accurate and contemporaneous UV and $\gamma$-ray observations, two alternative scenarios (Case A and Case B) were proposed,
with the Doppler factor fixed at 50 for modelling all the SEDs.
Figure \ref{figs:sed1} clearly shows that emission from a single zone is sufficient to account for the MWL SED of PKS 2155-304 across the three observations.
However, it should be noted that, in both cases, the electrons responsible for X-ray emission must be primarily governed by the synchrotron cooling process, with stochastic acceleration playing a negligible role.

Comparison between the theoretical predictions and the measured Fourier X-ray time lags across various sub-bands is shown in Figs. \ref{figs:ftlag1}, \ref{figs:ftlag2}, and \ref{figs:ftlag3}.
The predicted lag-frequency spectra are represented by the dotted and dashed lines, corresponding to Cases A and B, respectively.
From the figures, we find that all the predicted lag-frequency spectra in Case A do not match the results obtained from the three observations.
Meanwhile, we note that all the predicted lag-frequency spectra in Case B are reconciled with the measured results except for ObsID 0124930601.

By utilising a simple causality argument, we obtain the minimum variability timescales $\tau_{\rm var,m} \equiv  R' (1+z)/c\delta_{\rm D} \simeq 13$ and $10$ h 
for ObsIDs 0124930301 and 0124930501, respectively.
These may not be consistent with the measurable minimum variability timescales derived from the PSD analyses.
For  the two observations our PSD analyses  give the flux-weighted timescales\footnote{The variability timescale, corresponding to the frequency threshold $\nu_{\rm Pois}$, is weighted by the mean flux at four sub-bands.}
of $\tau_{\rm var,w} \sim 21$ and $32$ minutes,
which are comparable to the minimum two-point flux variability timescale reported in \cite{Bhatta2025ApJ}.
Thus, it is challenging to reproduce the rapid variabilities observed in the two observations.

Additionally, the time lags for the given values of the model parameters can be approximately estimated by means of the analytical relationship 
\begin{equation}
\tau_{\rm soft} \simeq - 26.8(1+z)^{1/2}\delta_{\rm D}^{-1/2}B'^{-3/2} \left(E_{\rm s,keV}^{-1/2} - E_{\rm h,keV}^{-1/2}\right)~\rm minutes
\end{equation}
in the case of the pure synchrotron cooling \citep{Finke2014ApJ}. Here, $E_{\rm s/h, keV}$ is the energy of the observed photons expressed in keV.
For ObsIDs 0124930301 and 0124930501, we can immediately derive $\tau_{0.5-2/2-10}\simeq 6$ hours with the parameter values in Case B,
and using the channel-centre  energies at soft and hard bands.
 This large lag is caused by the very small magnetic field strength adopted in the modelling.
Obviously, such a large lag between the main soft and hard bands is unrealistic (see Figs. \ref{figs:lc1},\ref{figs:lc2}, and \ref{figs:lc3}).
For Case A the resulting time lags between the various bands are comparable to the low-frequency Fourier time lags obtained with a purely numerical approach,\footnote{The time lag profiles in the Fourier frequency domain depend not only  on $B'$ and $\delta_{\rm D}$,
but also on the spectral shape of the injected electrons and diffusion coefficient \citep[for details see][]{Hu2024ApJ}.}
and are shown in Figs. \ref{figs:ftlag1}, \ref{figs:ftlag2}, and \ref{figs:ftlag3}.

Therefore, our results show that the one-zone SSC scenario faces difficulties in simultaneously accounting for the observed broad-band X-ray SED and the  inter-band time lags
in the Fourier-frequency domain.

In addition, the classical one-zone model is also challenged by the variability observed on various timescales.
On the one hand, various authors focus on the determination of the statistical properties of the variability on IDV timescales \citep[e.g.][]{Gaur2010ApJ,Zhongli2021ApJ,Bhatta2025ApJ}. 
The intra-day X-ray PSD analysis of PKS 2155-304 reveals a variable power-law PSD with indices spanning a surprisingly wide range, from $\lesssim1$ to $\gtrsim3$,
though it predominantly falls within the red noise regime.
The extensive range of indices suggests a scenario involving multiple turbulent regions within relativistic jets.

On the other hand, the PSD analyses of LTV and STV for PKS 2155-304 indicate distinct variability behaviours at different frequencies.
Variability at the IC-dominated frequencies (GeV and TeV $\gamma$-rays) can be characterised by flicker and/or pink-noise processes on  timescales from years to months,
whereas at synchrotron-dominated frequencies (Radio and Optical bands) damped and/or red-noise processes dominate on  timescales from years to weeks \citep{Goyal2020MNRAS}.
The X-ray variability can be represented by a power-law index of $\alpha\sim1$ extending to the variability timescales of $\sim$100 days.
The observed variabilities in different wavelengths are incompatible with the theoretical prediction of analytical studies \citep{Finke2014ApJ, Finke2015ApJ} 
 and numerical simulations \citep{Thiersen2022ApJ} within one-zone leptonic models.
 In such models, one should expect the same slopes of the variability PSDs across all the wavelengths 
because both low-energy synchrotron and high-energy IC emission are attributed to the same electron population.
The complicated variability may be a natural result of a structured or stratified jet that involves more than one emission zone. 

The PSD analysis over the probed timescales indicates that the blazar variabilities are driven by energy dissipation on widely different spatial scales.
It reveals a highly inhomogeneous jet, with structures present on both  large and small spatial scales, associated with different physical processes.

Furthermore, multi-zone emission models have also had success in modelling the observed polarisation of blazars \citep[see][and references therein]{Peceur2020MNRAS}.
Very recently, \cite{Kouch2024A&A} reported that the X-ray polarisation degree ($\Pi_{\rm X}\sim15\%-31\%$) of PKS 2155-304 observed with the Imaging X-ray Polarimetry Explorer (IXPE) 
is several times greater than the optical polarisation degree ($\Pi_{\rm O}$), along with the lack of temporal correlation between $\Pi_{\rm X}$ and $\Pi_{\rm O}$.
Meanwhile, the X-ray polarisation angle  ($\Psi_{\rm X}\sim125^\circ-129^\circ$) remained relatively stable,
whereas the optical polarisation angle  ($\Psi_{\rm O}$) exhibited achromatic variations during the IXPE pointing.
The findings indicate that the energetic electrons radiating in the X-ray regime are associated with energy dissipation through shocks rather than magnetic reconnection,
and reveal that the X-ray and optical emission come, at least partly, from different regions of the jet. 

In the shock acceleration scenario a fraction of the jet's kinetic power is dissipated, accelerating particles up to ultra-relativistic energies and producing the observed emission.
The formation of shocks requires that the jet is dominated by kinetic energy flux at the dissipation distance.
This implies that the energy density of the radiating particles is significantly larger than that of the magnetic field at the emitting region, where intense energy dissipation has occurred.
In this study, the kinetic-to-magnetic energy ratio $U'_{\rm e}/U'_{\rm B}$ at the emitting region is derived by jointly modelling the observed broad-band X-ray SED and inter-band time lags in the Fourier domain.
Based on a multi-zone leptonic model, we find that the ratio of energy densities $U'_{\rm e}/U'_{\rm B}$ is in the range $\sim$35--150 (see Table \ref{tabs:twozones}),
which agrees with the prediction of shocks acceleration. 
Additionally, the inferred power-law exponent of energetic electrons spectra $p\sim2.3-2.5$ is also compatible with the theoretical predictions of the shock acceleration \citep[e.g.][]{Kirk1996A&A,Sironi2009ApJ,Summerlin2012ApJ,Sironi2013ApJ}.
Furthermore, if the cooling is dominated by synchrotron emission, the maximum energy of electrons accelerated by quasi-parallel shocks is given by \citep[e.g.][]{Drury1983RPPh,Protheroe2004PASA,Rieger2007Ap&SS}
\begin{equation}
\gamma'_{\rm max}\simeq1.43\times10^6\frac{u'_{\rm sh}}{c}\left(\frac{\eta_{\rm g}}{10^4}\right)^{-1/2}\left(\frac{B'}{0.1}\right)^{-1/2},
\end{equation}
where $u_{\rm sh}'$ is the shock speed, and $\eta_{\rm g}>1$ is the gyrofactor reflecting electron acceleration efficiency.
Given $\gamma'_{\rm max}\sim(2-3)\times10^5$ inferred from our modelling, we estimate the gyrofactor $\eta_{\rm g}\gtrsim(5-7)\times10^4$, 
considering a steep spectral index ($p>2$) and adopting a mildly relativistic shock speed ($u'_{\rm sh}\gtrsim0.3c$) along with a typical magnetic field strength of $B'=0.1$ G.
This high $\eta_{\rm g}$ value is consistent with the results derived from the spectral fits of 13 TeV HBLs \citep{Inoue2016ApJ} and numerical experiments of shock acceleration \citep{Summerlin2012ApJ,Baring2017MNRAS}.

Combined with the satisfactory modelling of the observed X-ray LCs by varying the injection rate of the electrons, 
we conclude that the X-ray IDV of PKS 2155-304 is caused by the injection of energetic electrons, accelerated by shocks formed in a weakly magnetised jet.

\begin{acknowledgements}
We thank the editor and the anonymous referee for constructive comments that helped improve and clarify the paper.
This work is supported by the National Natural Science Foundation of China (NSFC-12263003, 12033006 and 12403048).
We are thankful to Mou-yuan Sun (XMU) and Zhen-yi Cai (USTC) for discussion on the PSD analysis. 
WH thanks Dan Wilkins for using GP.    
WH acknowledges support from the PhD Starting Fund program of JingGangShan University under Grant No. 2017KFW001.
\end{acknowledgements}

%
%

\bibliographystyle{aa}
\bibliography{sample} 

\begin{thebibliography}{101}
\expandafter\ifx\csname natexlab\endcsname\relax\def\natexlab#1{#1}\fi

\bibitem[{{Abdalla} {et~al.}(2020){Abdalla}, {Adam}, {Aharonian}, {Ait
  Benkhali}, {Ang{\"u}ner}, {Arakawa}, {Arcaro}, {Armand}, {Ashkar}, {Backes},
  {Barbosa Martins}, {Barnard}, {Becherini}, {Berge}, {Bernl{\"o}hr},
  {Blackwell}, {B{\"o}ttcher}, {Boisson}, {Bolmont}, {Bonnefoy}, {Bregeon},
  {Breuhaus}, {Brun}, {Brun}, {Bryan}, {B{\"u}chele}, {Bulik}, {Bylund},
  {Caroff}, {Carosi}, {Casanova}, {Cerruti}, {Chand}, {Chandra}, {Chen},
  {Colafrancesco}, {Cury{\l}o}, {Davids}, {Deil}, {Devin}, {deWilt}, {Dirson},
  {Djannati-Ata{\"\i}}, {Dmytriiev}, {Donath}, {Doroshenko}, {Dyks}, {Egberts},
  {Emery}, {Ernenwein}, {Eschbach}, {Feijen}, {Fegan}, {Fiasson}, {Fontaine},
  {Funk}, {F{\"u}{\ss}ling}, {Gabici}, {Gallant}, {Gat{\'e}}, {Giavitto},
  {Giunti}, {Glawion}, {Glicenstein}, {Gottschall}, {Grondin}, {Hahn}, {Haupt},
  {Heinzelmann}, {Henri}, {Hermann}, {Hinton}, {Hofmann}, {Hoischen}, {Holch},
  {Holler}, {Horns}, {Huber}, {Iwasaki}, {Jamrozy}, {Jankowsky}, {Jankowsky},
  {Jardin-Blicq}, {Jung-Richardt}, {Kastendieck}, {Katarzy{\'n}ski},
  {Katsuragawa}, {Katz}, {Khangulyan}, {Kh{\'e}lifi}, {King}, {Klepser},
  {Klu{\'z}niak}, {Komin}, {Kosack}, {Kostunin}, {Kreter}, {Lamanna},
  {Lemi{\`e}re}, {Lemoine-Goumard}, {Lenain}, {Leser}, {Levy}, {Lohse},
  {Lypova}, {Mackey}, {Majumdar}, {Malyshev}, {Marandon}, {Marcowith}, {Mares},
  {Mariaud}, {Mart{\'\i}-Devesa}, {Marx}, {Maurin}, {Meintjes}, {Mitchell},
  {Moderski}, {Mohamed}, {Mohrmann}, {Moore}, {Moulin}, {Muller}, {Murach},
  {Nakashima}, {de Naurois}, {Ndiyavala}, {Niederwanger}, {Niemiec}, {Oakes},
  {O'Brien}, {Odaka}, {Ohm}, {de Ona Wilhelmi}, {Ostrowski}, {Oya}, {Panter},
  {Parsons}, {Perennes}, {Petrucci}, {Peyaud}, {Piel}, {Pita}, {Poireau},
  {Priyana Noel}, {Prokhorov}, {Prokoph}, {P{\"u}hlhofer}, {Punch},
  {Quirrenbach}, {Raab}, {Rauth}, {Reimer}, {Reimer}, {Remy}, {Renaud},
  {Rieger}, {Rinchiuso}, {Romoli}, {Rowell}, {Rudak}, {Ruiz-Velasco},
  {Sahakian}, {Sailer}, {Saito}, {Sanchez}, {Santangelo}, {Sasaki},
  {Schlickeiser}, {Sch{\"u}ssler}, {Schulz}, {Schutte}, {Schwanke},
  {Schwemmer}, {Seglar-Arroyo}, {Senniappan}, {Seyffert}, {Shafi},
  {Shiningayamwe}, {Simoni}, {Sinha}, {Sol}, {Specovius}, {Spir-Jacob},
  {Stawarz}, {Steenkamp}, {Stegmann}, {Steppa}, {Takahashi}, {Tavernier},
  {Taylor}, {Terrier}, {Tiziani}, {Tluczykont}, {Trichard}, {Tsirou}, {Tsuji},
  {Tuffs}, {Uchiyama}, {van der Walt}, {van Eldik}, {van Rensburg}, {van
  Soelen}, {Vasileiadis}, {Veh}, {Venter}, {Vincent}, {Vink}, {V{\"o}lk},
  {Vuillaume}, {Wadiasingh}, {Wagner}, {White}, {Wierzcholska}, {Yang},
  {Yoneda}, {Zacharias}, {Zanin}, {Zdziarski}, {Zech}, {Zorn}, {{\.Z}ywucka},
  {Madejski}, {Nalewajko}, {Madsen}, {Chiang}, {Balokovi{\'c}}, {Paneque},
  {Furniss}, {Hayashida}, {Urry}, {Ajello}, {Harrison}, {Giebels}, {Stern},
  {Forster}, {Giommi}, {Perri}, {Puccetti}, {Zoglauer}, \&
  {Tagliaferri}}]{Abdalla2020A&A}
{Abdalla}, H., {Adam}, R., {Aharonian}, F., {et~al.} 2020, \aap, 639, A42

\bibitem[{{Abdo} {et~al.}(2011){Abdo}, {Ackermann}, {Ajello}, {Allafort},
  {Baldini}, {Ballet}, {Barbiellini}, {Baring}, {Bastieri}, {Bechtol},
  {Bellazzini}, {Berenji}, {Blandford}, {Bloom}, {Bonamente}, {Borgland},
  {Bouvier}, {Brandt}, {Bregeon}, {Brez}, {Brigida}, {Bruel}, {Buehler},
  {Buson}, {Caliandro}, {Cameron}, {Cannon}, {Caraveo}, {Carrigan},
  {Casandjian}, {Cavazzuti}, {Cecchi}, {{\c{C}}elik}, {Charles}, {Chekhtman},
  {Cheung}, {Chiang}, {Ciprini}, {Claus}, {Cohen-Tanugi}, {Conrad}, {Cutini},
  {Dermer}, {de Palma}, {Silva}, {Drell}, {Dubois}, {Dumora}, {Favuzzi},
  {Fegan}, {Ferrara}, {Focke}, {Fortin}, {Frailis}, {Fuhrmann}, {Fukazawa},
  {Funk}, {Fusco}, {Gargano}, {Gasparrini}, {Gehrels}, {Germani}, {Giglietto},
  {Giordano}, {Giroletti}, {Glanzman}, {Godfrey}, {Grenier}, {Guillemot},
  {Guiriec}, {Hayashida}, {Hays}, {Horan}, {Hughes}, {J{\'o}hannesson},
  {Johnson}, {Johnson}, {Kadler}, {Kamae}, {Katagiri}, {Kataoka},
  {Kn{\"o}dlseder}, {Kuss}, {Lande}, {Latronico}, {Lee}, {Lemoine-Goumard},
  {Longo}, {Loparco}, {Lott}, {Lovellette}, {Lubrano}, {Madejski}, {Makeev},
  {Max-Moerbeck}, {Mazziotta}, {McEnery}, {Mehault}, {Michelson},
  {Mitthumsiri}, {Mizuno}, {Moiseev}, {Monte}, {Monzani}, {Morselli},
  {Moskalenko}, {Murgia}, {Naumann-Godo}, {Nishino}, {Nolan}, {Norris}, {Nuss},
  {Ohsugi}, {Okumura}, {Omodei}, {Orlando}, {Ormes}, {Paneque}, {Panetta},
  {Parent}, {Pavlidou}, {Pearson}, {Pelassa}, {Pepe}, {Pesce-Rollins}, {Piron},
  {Porter}, {Rain{\`o}}, {Rando}, {Razzano}, {Readhead}, {Reimer}, {Reimer},
  {Richards}, {Ripken}, {Ritz}, {Roth}, {Sadrozinski}, {Sanchez}, {Sander},
  {Scargle}, {Sgr{\`o}}, {Siskind}, {Smith}, {Spandre}, {Spinelli}, {Stawarz},
  {Stevenson}, {Strickman}, {Sokolovsky}, {Suson}, {Takahashi}, {Takahashi},
  {Tanaka}, {Thayer}, {Thayer}, {Thompson}, {Tibaldo}, {Torres}, {Tosti},
  {Tramacere}, {Uchiyama}, {Usher}, {Vandenbroucke}, {Vasileiou}, {Vilchez},
  {Vitale}, {Waite}, {Wang}, {Wehrle}, {Winer}, {Wood}, {Yang}, {Ylinen},
  {Zensus}, {Ziegler}, {Fermi LAT Collaboration}, {Aleksi{\'c}}, {Antonelli},
  {Antoranz}, {Backes}, {Barrio}, {Becerra Gonz{\'a}lez}, {Bednarek},
  {Berdyugin}, {Berger}, {Bernardini}, {Biland}, {Blanch}, {Bock}, {Boller},
  {Bonnoli}, {Bordas}, {Borla Tridon}, {Bosch-Ramon}, {Bose}, {Braun}, {Bretz},
  {Camara}, {Carmona}, {Carosi}, {Colin}, {Colombo}, {Contreras}, {Cortina},
  {Covino}, {Dazzi}, {de Angelis}, {De Cea del Pozo}, {De Lotto}, {De Maria},
  {De Sabata}, {Delgado Mendez}, {Diago Ortega}, {Doert}, {Dom{\'\i}nguez},
  {Dominis Prester}, {Dorner}, {Doro}, {Elsaesser}, {Ferenc}, {Fonseca},
  {Font}, {Garc{\'\i}a L{\'o}pez}, {Garczarczyk}, {Gaug}, {Giavitto},
  {Godinovi}, {Hadasch}, {Herrero}, {Hildebrand}, {H{\"o}hne-M{\"o}nch},
  {Hose}, {Hrupec}, {Jogler}, {Klepser}, {Kr{\"a}henb{\"u}hl}, {Kranich},
  {Krause}, {La Barbera}, {Leonardo}, {Lindfors}, {Lombardi}, {L{\'o}pez},
  {Lorenz}, {Majumdar}, {Makariev}, {Maneva}, {Mankuzhiyil}, {Mannheim},
  {Maraschi}, {Mariotti}, {Mart{\'\i}nez}, {Mazin}, {Meucci}, {Miranda},
  {Mirzoyan}, {Miyamoto}, {Mold{\'o}n}, {Moralejo}, {Nieto}, {Nilsson},
  {Orito}, {Oya}, {Paoletti}, {Paredes}, {Partini}, {Pasanen}, {Pauss},
  {Pegna}, {Perez-Torres}, {Persic}, {Peruzzo}, {Pochon}, {Prada Moroni},
  {Prada}, {Prandini}, {Puchades}, {Puljak}, {Reichardt}, {Reinthal}, {Rhode},
  {Rib{\'o}}, {Rico}, {Rissi}, {R{\"u}gamer}, {Saggion}, {Saito}, {Saito},
  {Salvati}, {S{\'a}nchez-Conde}, {Satalecka}, {Scalzotto}, {Scapin},
  {Schultz}, {Schweizer}, {Shayduk}, {Shore}, {Sierpowska-Bartosik},
  {Sillanp{\"a}{\"a}}, {Sitarek}, {Sobczynska}, {Spanier}, {Spiro}, {Stamerra},
  {Steinke}, {Storz}, {Strah}, {Struebig}, {Suric}, {Takalo}, {Tavecchio},
  {Temnikov}, {Terzi{\'c}}, {Tescaro}, {Teshima}, {Vankov}, {Wagner},
  {Weitzel}, {Zabalza}, {Zandanel}, {Zanin}, {MAGIC Collaboration}, {Acciari},
  {Arlen}, {Aune}, {Benbow}, {Boltuch}, {Bradbury}, {Buckley}, {Bugaev},
  {Cannon}, {Cesarini}, {Ciupik}, {Cui}, {Dickherber}, {Errando}, {Falcone},
  {Finley}, {Finnegan}, {Fortson}, {Furniss}, {Galante}, {Gall}, {Gillanders},
  {Godambe}, {Grube}, {Guenette}, {Gyuk}, {Hanna}, {Holder}, {Huang}, {Hui},
  {Humensky}, {Kaaret}, {Karlsson}, {Kertzman}, {Kieda}, {Konopelko},
  {Krawczynski}, {Krennrich}, {Lang}, {Maier}, {McArthur}, {McCann},
  {McCutcheon}, {Moriarty}, {Mukherjee}, {Ong}, {Otte}, {Pandel}, {Perkins},
  {Pichel}, {Pohl}, {Quinn}, {Ragan}, {Reyes}, {Reynolds}, {Roache}, {Rose},
  {Rovero}, {Schroedter}, {Sembroski}, {Senturk}, {Steele}, {Swordy},
  {Te{\v{s}}i{\'c}}, {Theiling}, {Thibadeau}, {Varlotta}, {Vincent}, {Wakely},
  {Ward}, {Weekes}, {Weinstein}, {Weisgarber}, {Williams}, {Wood}, {Zitzer},
  {VERITAS Collaboration}, {Villata}, {Raiteri}, {Aller}, {Aller}, {Arkharov},
  {Blinov}, {Calcidese}, {Chen}, {Efimova}, {Kimeridze}, {Konstantinova},
  {Kopatskaya}, {Koptelova}, {Kurtanidze}, {Kurtanidze}, {L{\"a}hteenm{\"a}ki},
  {Larionov}, {Larionova}, {Larionova}, {Ligustri}, {Morozova}, {Nikolashvili},
  {Sigua}, {Troitsky}, {Angelakis}, {Capalbi}, {Carrami{\~n}ana}, {Carrasco},
  {Cassaro}, {de la Fuente}, {Gurwell}, {Kovalev}, {Kovalev}, {Krichbaum},
  {Krimm}, {Leto}, {Lister}, {Maccaferri}, {Moody}, {Mori}, {Nestoras},
  {Orlati}, {Pagani}, {Pace}, {Pearson}, {Perri}, {Piner}, {Pushkarev}, {Ros},
  {Sadun}, {Sakamoto}, {Tornikoski}, {Yatsu}, \& {Zook}}]{Abdo2011ApJ501}
{Abdo}, A.~A., {Ackermann}, M., {Ajello}, M., {et~al.} 2011, \apj, 727, 129

\bibitem[{{Aguilar-Ruiz} {et~al.}(2022){Aguilar-Ruiz}, {Fraija},
  {Galv{\'a}n-G{\'a}mez}, \& {Ben{\'\i}tez}}]{Aguilar2022MNRAS}
{Aguilar-Ruiz}, E., {Fraija}, N., {Galv{\'a}n-G{\'a}mez}, A., \&
  {Ben{\'\i}tez}, E. 2022, \mnras, 512, 1557

\bibitem[{{Aharonian} {et~al.}(2009{\natexlab{a}}){Aharonian}, {Akhperjanian},
  {Anton}, {Barres de Almeida}, {Bazer-Bachi}, {Becherini}, {Behera}, {Benbow},
  {Bernl{\"o}hr}, {Boisson}, {Bochow}, {Borrel}, {Brion}, {Brucker}, {Brun},
  {B{\"u}hler}, {Bulik}, {B{\"u}sching}, {Boutelier}, {Chadwick},
  {Charbonnier}, {Chaves}, {Cheesebrough}, {Chounet}, {Clapson}, {Coignet},
  {Costamante}, {Dalton}, {Daniel}, {Davids}, {Degrange}, {Deil}, {Dickinson},
  {Djannati-Ata{\"\i}}, {Domainko}, {O'C. Drury}, {Dubois}, {Dubus}, {Dyks},
  {Dyrda}, {Egberts}, {Emmanoulopoulos}, {Espigat}, {Farnier}, {Feinstein},
  {Fiasson}, {F{\"o}rster}, {Fontaine}, {F{\"u}{\ss}ling}, {Gabici}, {Gallant},
  {G{\'e}rard}, {Giebels}, {Glicenstein}, {Gl{\"u}ck}, {Goret}, {G{\"o}hring},
  {Hauser}, {Hauser}, {Heinz}, {Heinzelmann}, {Henri}, {Hermann}, {Hinton},
  {Hoffmann}, {Hofmann}, {Holleran}, {Hoppe}, {Horns}, {Jacholkowska}, {de
  Jager}, {Jahn}, {Jung}, {Katarzy{\'n}ski}, {Katz}, {Kaufmann}, {Kendziorra},
  {Kerschhaggl}, {Khangulyan}, {Kh{\'e}lifi}, {Keogh}, {Klu{\'z}niak},
  {Kneiske}, {Komin}, {Kosack}, {Lamanna}, {Lenain}, {Lohse}, {Marandon},
  {Martin}, {Martineau-Huynh}, {Marcowith}, {Maurin}, {McComb}, {Medina},
  {Moderski}, {Monard}, {Moulin}, {Naumann-Godo}, {de Naurois}, {Nedbal},
  {Nekrassov}, {Niemiec}, {Nolan}, {Ohm}, {Olive}, {de O{\~n}a Wilhelmi},
  {Orford}, {Ostrowski}, {Panter}, {Paz Arribas}, {Pedaletti}, {Pelletier},
  {Petrucci}, {Pita}, {P{\"u}hlhofer}, {Punch}, {Quirrenbach}, {Raubenheimer},
  {Raue}, {Rayner}, {Renaud}, {Rieger}, {Ripken}, {Rob}, {Rosier-Lees},
  {Rowell}, {Rudak}, {Rulten}, {Ruppel}, {Sahakian}, {Santangelo},
  {Schlickeiser}, {Sch{\"o}ck}, {Schr{\"o}der}, {Schwanke}, {Schwarzburg},
  {Schwemmer}, {Shalchi}, {Sikora}, {Skilton}, {Sol}, {Spangler}, {Stawarz},
  {Steenkamp}, {Stegmann}, {Superina}, {Szostek}, {Tam}, {Tavernet}, {Terrier},
  {Tibolla}, {Tluczykont}, {van Eldik}, {Vasileiadis}, {Venter}, {Venter},
  {Vialle}, {Vincent}, {Vivier}, {V{\"o}lk}, {Volpe}, {Wagner}, {Ward},
  {Zdziarski}, \& {Zech}}]{Aharonian2009A&A}
{Aharonian}, F., {Akhperjanian}, A.~G., {Anton}, G., {et~al.}
  2009{\natexlab{a}}, \aap, 502, 749

\bibitem[{{Aharonian} {et~al.}(2009{\natexlab{b}}){Aharonian}, {Akhperjanian},
  {Anton}, {Barres de Almeida}, {Bazer-Bachi}, {Becherini}, {Behera},
  {Bernl{\"o}hr}, {Boisson}, {Bochow}, {Borrel}, {Brion}, {Brucker}, {Brun},
  {B{\"u}hler}, {Bulik}, {B{\"u}sching}, {Boutelier}, {Chadwick},
  {Charbonnier}, {Chaves}, {Cheesebrough}, {Chounet}, {Clapson}, {Coignet},
  {Dalton}, {Daniel}, {Davids}, {Degrange}, {Deil}, {Dickinson},
  {Djannati-Ata{\"\i}}, {Domainko}, {O'C. Drury}, {Dubois}, {Dubus}, {Dyks},
  {Dyrda}, {Egberts}, {Emmanoulopoulos}, {Espigat}, {Farnier}, {Feinstein},
  {Fiasson}, {F{\"o}rster}, {Fontaine}, {F{\"u}{\ss}ling}, {Gabici}, {Gallant},
  {G{\'e}rard}, {Giebels}, {Glicenstein}, {Gl{\"u}ck}, {Goret}, {G{\"o}hring},
  {Hauser}, {Hauser}, {Heinz}, {Heinzelmann}, {Henri}, {Hermann}, {Hinton},
  {Hoffmann}, {Hofmann}, {Holleran}, {Hoppe}, {Horns}, {Jacholkowska}, {de
  Jager}, {Jahn}, {Jung}, {Katarzy{\'n}ski}, {Katz}, {Kaufmann}, {Kendziorra},
  {Kerschhaggl}, {Khangulyan}, {Kh{\'e}lifi}, {Keogh}, {Klu{\'z}niak}, {Komin},
  {Kosack}, {Lamanna}, {Lenain}, {Lohse}, {Marandon}, {Martin},
  {Martineau-Huynh}, {Marcowith}, {Maurin}, {McComb}, {Medina}, {Moderski},
  {Moulin}, {Naumann-Godo}, {de Naurois}, {Nedbal}, {Nekrassov}, {Niemiec},
  {Nolan}, {Ohm}, {Olive}, {de O{\~n}a Wilhelmi}, {Orford}, {Ostrowski},
  {Panter}, {Arribas}, {Pedaletti}, {Pelletier}, {Petrucci}, {Pita},
  {P{\"u}hlhofer}, {Punch}, {Quirrenbach}, {Raubenheimer}, {Raue}, {Rayner},
  {Renaud}, {Rieger}, {Ripken}, {Rob}, {Rosier-Lees}, {Rowell}, {Rudak},
  {Rulten}, {Ruppel}, {Sahakian}, {Santangelo}, {Schlickeiser}, {Sch{\"o}ck},
  {Schr{\"o}der}, {Schwanke}, {Schwarzburg}, {Schwemmer}, {Shalchi}, {Sikora},
  {Skilton}, {Sol}, {Spangler}, {Stawarz}, {Steenkamp}, {Stegmann}, {Superina},
  {Szostek}, {Tam}, {Tavernet}, {Terrier}, {Tibolla}, {van Eldik},
  {Vasileiadis}, {Venter}, {Venter}, {Vialle}, {Vincent}, {Vivier}, {V{\"o}lk},
  {Volpe}, {Wagner}, {Ward}, {Zdziarski}, {Zech}, {H.~E.~S.~S. Collaboration},
  {Abdo}, {Ackermann}, {Ajello}, {Atwood}, {Axelsson}, {Baldini}, {Ballet},
  {Barbiellini}, {Baring}, {Bastieri}, {Battelino}, {Baughman}, {Bechtol},
  {Bellazzini}, {Berenji}, {Bloom}, {Bonamente}, {Borgland}, {Bregeon}, {Brez},
  {Brigida}, {Bruel}, {Caliandro}, {Cameron}, {Caraveo}, {Casandjian},
  {Cavazzuti}, {Cecchi}, {Charles}, {Chekhtman}, {Chen}, {Cheung}, {Chiang},
  {Ciprini}, {Claus}, {Cohen-Tanugi}, {Colafrancesco}, {Conrad}, {Costamante},
  {Cutini}, {Dermer}, {de Angelis}, {de Palma}, {Digel}, {do Couto e Silva},
  {Drell}, {Dubois}, {Dubus}, {Dumora}, {Farnier}, {Favuzzi}, {Fegan},
  {Ferrara}, {Fleury}, {Focke}, {Frailis}, {Fukazawa}, {Funk}, {Fusco},
  {Gargano}, {Gasparrini}, {Gehrels}, {Germani}, {Giebels}, {Giglietto},
  {Giordano}, {Grondin}, {Grove}, {Guillemot}, {Guiriec}, {Hanabata},
  {Harding}, {Hayashida}, {Hays}, {Horan}, {J{\'o}hannesson}, {Johnson},
  {Johnson}, {Johnson}, {Kadler}, {Kamae}, {Katagiri}, {Kataoka}, {Kerr},
  {Kn{\"o}dlseder}, {Kuehn}, {Kuss}, {Lande}, {Latronico}, {Lee},
  {Lemoine-Goumard}, {Longo}, {Loparco}, {Lott}, {Lovellette}, {Madejski},
  {Makeev}, {Mazziotta}, {McEnery}, {Meurer}, {Michelson}, {Mitthumsiri},
  {Mizuno}, {Moiseev}, {Monte}, {Monzani}, {Morselli}, {Moskalenko}, {Murgia},
  {Nolan}, {Nuss}, {Ohsugi}, {Omodei}, {Orlando}, {Ormes}, {Paneque},
  {Panetta}, {Parent}, {Pelassa}, {Pepe}, {Pesce-Rollins}, {Piron}, {Porter},
  {Rain{\`o}}, {Razzano}, {Reimer}, {Reimer}, {Reposeur}, {Ritz}, {Rodriguez},
  {Ryde}, {Sadrozinski}, {Sanchez}, {Sander}, {Scargle}, {Schalk},
  {Sellerholm}, {Sgr{\`o}}, {Shaw}, {Smith}, {Spandre}, {Spinelli}, {Starck},
  {Strickman}, {Tajima}, {Takahashi}, {Takahashi}, {Tanaka}, {Thayer},
  {Thompson}, {Tibaldo}, {Torres}, {Tosti}, {Tramacere}, {Uchiyama}, {Usher},
  {Vilchez}, {Villata}, {Vitale}, {Waite}, {Wood}, {Ylinen}, {Ziegler}, \&
  {Fermi LAT Collaboration}}]{Aharonian2009ApJ}
{Aharonian}, F., {Akhperjanian}, A.~G., {Anton}, G., {et~al.}
  2009{\natexlab{b}}, \apjl, 696, L150

\bibitem[{{Aharonian} {et~al.}(2007){Aharonian}, {Akhperjanian}, {Bazer-Bachi},
  {Behera}, {Beilicke}, {Benbow}, {Berge}, {Bernl{\"o}hr}, {Boisson}, {Bolz},
  {Borrel}, {Boutelier}, {Braun}, {Brion}, {Brown}, {B{\"u}hler},
  {B{\"u}sching}, {Bulik}, {Carrigan}, {Chadwick}, {Clapson}, {Chounet},
  {Coignet}, {Cornils}, {Costamante}, {Degrange}, {Dickinson},
  {Djannati-Ata{\"\i}}, {Domainko}, {Drury}, {Dubus}, {Dyks}, {Egberts},
  {Emmanoulopoulos}, {Espigat}, {Farnier}, {Feinstein}, {Fiasson},
  {F{\"o}rster}, {Fontaine}, {Funk}, {Funk}, {F{\"u}{\ss}ling}, {Gallant},
  {Giebels}, {Glicenstein}, {Gl{\"u}ck}, {Goret}, {Hadjichristidis}, {Hauser},
  {Hauser}, {Heinzelmann}, {Henri}, {Hermann}, {Hinton}, {Hoffmann}, {Hofmann},
  {Holleran}, {Hoppe}, {Horns}, {Jacholkowska}, {de Jager}, {Kendziorra},
  {Kerschhaggl}, {Kh{\'e}lifi}, {Komin}, {Kosack}, {Lamanna}, {Latham}, {Le
  Gallou}, {Lemi{\`e}re}, {Lemoine-Goumard}, {Lenain}, {Lohse}, {Martin},
  {Martineau-Huynh}, {Marcowith}, {Masterson}, {Maurin}, {McComb}, {Moderski},
  {Moulin}, {de Naurois}, {Nedbal}, {Nolan}, {Olive}, {Orford}, {Osborne},
  {Ostrowski}, {Panter}, {Pedaletti}, {Pelletier}, {Petrucci}, {Pita},
  {P{\"u}hlhofer}, {Punch}, {Ranchon}, {Raubenheimer}, {Raue}, {Rayner},
  {Renaud}, {Ripken}, {Rob}, {Rolland}, {Rosier-Lees}, {Rowell}, {Rudak},
  {Ruppel}, {Sahakian}, {Santangelo}, {Saug{\'e}}, {Schlenker}, {Schlickeiser},
  {Schr{\"o}der}, {Schwanke}, {Schwarzburg}, {Schwemmer}, {Shalchi}, {Sol},
  {Spangler}, {Stawarz}, {Steenkamp}, {Stegmann}, {Superina}, {Tam},
  {Tavernet}, {Terrier}, {van Eldik}, {Vasileiadis}, {Venter}, {Vialle},
  {Vincent}, {Vivier}, {V{\"o}lk}, {Volpe}, {Wagner}, {Ward}, \&
  {Zdziarski}}]{Aharonian2007ApJ}
{Aharonian}, F., {Akhperjanian}, A.~G., {Bazer-Bachi}, A.~R., {et~al.} 2007,
  \apjl, 664, L71

\bibitem[{{Arnaud}(1996)}]{Arnaud_1996}
{Arnaud}, K.~A. 1996, in Astronomical Society of the Pacific Conference Series,
  Vol. 101, Astronomical Data Analysis Software and Systems V, ed. G.~H.
  {Jacoby} \& J.~{Barnes}, 17

\bibitem[{{Baring} {et~al.}(2017){Baring}, {B{\"o}ttcher}, \&
  {Summerlin}}]{Baring2017MNRAS}
{Baring}, M.~G., {B{\"o}ttcher}, M., \& {Summerlin}, E.~J. 2017, \mnras, 464,
  4875

\bibitem[{{Begelman} {et~al.}(2008){Begelman}, {Fabian}, \&
  {Rees}}]{Begelman2008MNRAS}
{Begelman}, M.~C., {Fabian}, A.~C., \& {Rees}, M.~J. 2008, \mnras, 384, L19

\bibitem[{{Bhagwan} {et~al.}(2014){Bhagwan}, {Gupta}, {Papadakis}, \&
  {Wiita}}]{Bhagwan2014MNRAS}
{Bhagwan}, J., {Gupta}, A.~C., {Papadakis}, I.~E., \& {Wiita}, P.~J. 2014,
  \mnras, 444, 3647

\bibitem[{Bhagwana {et~al.}(2015)Bhagwana, Gupta, Papadakis, \&
  Wiita}]{Bhagwana2015FluxAS}
Bhagwana, J., Gupta, A.~C., Papadakis, I., \& Wiita, P.~J. 2015, New Astronomy,
  44, 21

\bibitem[{{Bhatta} {et~al.}(2025){Bhatta}, {Chaudhary}, {Dhital}, {Adhikari},
  {Mohorian}, {Dinesh}, {P{\'a}nis}, {Neupane}, \& {Maharjan}}]{Bhatta2025ApJ}
{Bhatta}, G., {Chaudhary}, S.~C., {Dhital}, N., {et~al.} 2025, \apj, 981, 118

\bibitem[{{Blandford} \& {K{\"o}nigl}(1979)}]{Blandford1979}
{Blandford}, R.~D. \& {K{\"o}nigl}, A. 1979, \apj, 232, 34

\bibitem[{{B{\"o}ttcher} {et~al.}(2013){B{\"o}ttcher}, {Reimer}, {Sweeney}, \&
  {Prakash}}]{Bottcher2013ApJ}
{B{\"o}ttcher}, M., {Reimer}, A., {Sweeney}, K., \& {Prakash}, A. 2013, \apj,
  768, 54

\bibitem[{{Boutelier} {et~al.}(2008){Boutelier}, {Henri}, \&
  {Petrucci}}]{Boutelier2008MNRAS}
{Boutelier}, T., {Henri}, G., \& {Petrucci}, P.~O. 2008, \mnras, 390, L73

\bibitem[{{Calafut} \& {Wiita}(2015)}]{Calafut2015JApA}
{Calafut}, V. \& {Wiita}, P.~J. 2015, Journal of Astrophysics and Astronomy,
  36, 255

\bibitem[{{Cao} {et~al.}(2020){Cao}, {Yang}, {Yang}, \& {Wang}}]{Cao2020PASJ}
{Cao}, G., {Yang}, C., {Yang}, J., \& {Wang}, J. 2020, \pasj, 72, 20

\bibitem[{{Cerutti} {et~al.}(2012){Cerutti}, {Uzdensky}, \&
  {Begelman}}]{Cerutti2012ApJ}
{Cerutti}, B., {Uzdensky}, D.~A., \& {Begelman}, M.~C. 2012, \apj, 746, 148

\bibitem[{{Chadwick} {et~al.}(1999){Chadwick}, {Lyons}, {McComb}, {Orford},
  {Osborne}, {Rayner}, {Shaw}, {Turver}, \& {Wieczorek}}]{Chadwick1999ApJ}
{Chadwick}, P.~M., {Lyons}, K., {McComb}, T.~J.~L., {et~al.} 1999, \apj, 513,
  161

\bibitem[{{Chakrabarti} \& {Wiita}(1993)}]{Chakrabarti1993ApJ}
{Chakrabarti}, S.~K. \& {Wiita}, P.~J. 1993, \apj, 411, 602

\bibitem[{{Chen}(2018)}]{Chen2018ApJS}
{Chen}, L. 2018, \apjs, 235, 39

\bibitem[{{Comisso} \& {Sironi}(2018)}]{Comisso2018PhRvL}
{Comisso}, L. \& {Sironi}, L. 2018, \prl, 121, 255101

\bibitem[{{Deng} {et~al.}(2021){Deng}, {Hu}, {Lu}, \& {Dai}}]{Deng2021MNRAS}
{Deng}, X.-C., {Hu}, W., {Lu}, F.-W., \& {Dai}, B.-Z. 2021, \mnras, 504, 878

\bibitem[{{Dermer} {et~al.}(2009){Dermer}, {Finke}, {Krug}, \&
  {B{\"o}ttcher}}]{Dermer2009ApJ}
{Dermer}, C.~D., {Finke}, J.~D., {Krug}, H., \& {B{\"o}ttcher}, M. 2009, \apj,
  692, 32

\bibitem[{{Devanand} {et~al.}(2022){Devanand}, {Gupta}, {Jithesh}, \&
  {Wiita}}]{Devanand2022ApJ}
{Devanand}, P.~U., {Gupta}, A.~C., {Jithesh}, V., \& {Wiita}, P.~J. 2022, \apj,
  939, 80

\bibitem[{{Drury}(1983)}]{Drury1983RPPh}
{Drury}, L.~O. 1983, Reports on Progress in Physics, 46, 973

\bibitem[{{Falomo} {et~al.}(1993){Falomo}, {Pesce}, \&
  {Treves}}]{Falomo1993ApJ}
{Falomo}, R., {Pesce}, J.~E., \& {Treves}, A. 1993, \apjl, 411, L63

\bibitem[{{Finke} \& {Becker}(2014)}]{Finke2014ApJ}
{Finke}, J.~D. \& {Becker}, P.~A. 2014, \apj, 791, 21

\bibitem[{{Finke} \& {Becker}(2015)}]{Finke2015ApJ}
{Finke}, J.~D. \& {Becker}, P.~A. 2015, \apj, 809, 85

\bibitem[{{Foreman-Mackey} {et~al.}(2017){Foreman-Mackey}, {Agol},
  {Ambikasaran}, \& {Angus}}]{Foreman2017AJ}
{Foreman-Mackey}, D., {Agol}, E., {Ambikasaran}, S., \& {Angus}, R. 2017, \aj,
  154, 220

\bibitem[{{Franceschini} {et~al.}(2008){Franceschini}, {Rodighiero}, \&
  {Vaccari}}]{Franceschini2008A&A}
{Franceschini}, A., {Rodighiero}, G., \& {Vaccari}, M. 2008, \aap, 487, 837

\bibitem[{{Gasparyan} {et~al.}(2022){Gasparyan}, {B{\'e}gu{\'e}}, \&
  {Sahakyan}}]{Gasparyan2022MNRAS}
{Gasparyan}, S., {B{\'e}gu{\'e}}, D., \& {Sahakyan}, N. 2022, \mnras, 509, 2102

\bibitem[{{Gaur} {et~al.}(2017){Gaur}, {Chen}, {Misra}, {Sahayanathan}, {Gu},
  {Kushwaha}, \& {Dewangan}}]{Gaur2017ApJ}
{Gaur}, H., {Chen}, L., {Misra}, R., {et~al.} 2017, \apj, 850, 209

\bibitem[{{Gaur} {et~al.}(2010){Gaur}, {Gupta}, {Lachowicz}, \&
  {Wiita}}]{Gaur2010ApJ}
{Gaur}, H., {Gupta}, A.~C., {Lachowicz}, P., \& {Wiita}, P.~J. 2010, \apj, 718,
  279

\bibitem[{{Ghisellini} \& {Tavecchio}(2009)}]{Ghisellini2009MNRAS}
{Ghisellini}, G. \& {Tavecchio}, F. 2009, \mnras, 397, 985

\bibitem[{{Ghisellini} {et~al.}(2005){Ghisellini}, {Tavecchio}, \&
  {Chiaberge}}]{Ghisellini2005A&A}
{Ghisellini}, G., {Tavecchio}, F., \& {Chiaberge}, M. 2005, \aap, 432, 401

\bibitem[{{Ghisellini} {et~al.}(2014){Ghisellini}, {Tavecchio}, {Maraschi},
  {Celotti}, \& {Sbarrato}}]{Ghisellini2014Natur}
{Ghisellini}, G., {Tavecchio}, F., {Maraschi}, L., {Celotti}, A., \&
  {Sbarrato}, T. 2014, \nat, 515, 376

\bibitem[{{Goyal}(2020)}]{Goyal2020MNRAS}
{Goyal}, A. 2020, \mnras, 494, 3432

\bibitem[{{H.~E.~S.~S. Collaboration} {et~al.}(2012){H.~E.~S.~S.
  Collaboration}, {Abramowski}, {Acero}, {Aharonian}, {Akhperjanian}, {Anton},
  {Balzer}, {Barnacka}, {Barres de Almeida}, {Becherini}, {Becker}, {Behera},
  {Benbow}, {Bernl{\"o}hr}, {Bochow}, {Boisson}, {Bolmont}, {Bordas},
  {Bouteilier}, {Brucker}, {Brun}, {Brun}, {Bulik}, {B{\"u}sching}, {Carrigan},
  {Casanova}, {Cerruti}, {Chadwick}, {Charbonnier}, {Chaves}, {Cheesebrough},
  {Chounet}, {Clapson}, {Coignet}, {Cologna}, {Colom}, {Conrad}, {Coudreau},
  {Dalton}, {Daniel}, {Davids}, {Degrange}, {Deil}, {Dickinson},
  {Djannati-Ata{\"\i}}, {Domainko}, {Drury}, {Dubois}, {Dubus}, {Dutson},
  {Dyks}, {Dyrda}, {Edwards}, {Egberts}, {Eger}, {Espigat}, {Fallon},
  {Farnier}, {Fegan}, {Feinstein}, {Fernandes}, {Fiasson}, {Fontaine},
  {F{\"o}rster}, {F{\"u}{\ss}ling}, {Gallant}, {Gast}, {Gaylard}, {G{\'e}rard},
  {Gerbig}, {Giebels}, {Glicenstein}, {Gl{\"u}ck}, {Goret}, {G{\"o}ring},
  {H{\"a}ffner}, {Hague}, {Hampf}, {Hauser}, {Heinz}, {Heinzelmann}, {Henri},
  {Hermann}, {Hinton}, {Hoffmann}, {Hofmann}, {Hofvergerg}, {Holler}, {Horns},
  {Jacholkowska}, {de Jager}, {Jahn}, {Jamrozy}, {Jung}, {Kastendieck},
  {Katarzy{\'n}ski}, {Katz}, {Kaufmann}, {Keogh}, {Khangulyan}, {Kh{\'e}lifi},
  {Klein}, {Klochkov}, {Klu{\'z}niak}, {Kneiske}, {Komin}, {Kosack},
  {Kossakowski}, {Kubanek}, {Laffon}, {Lamanna}, {Lennarz}, {Lenain}, {Lohse},
  {Lopatin}, {Lu}, {Marandon}, {Marcowith}, {Martin}, {Masbou}, {Maurin},
  {Maxted}, {McComb}, {Medina}, {M{\'e}hault}, {Melady}, {Nguyen}, {Moderski},
  {Monard}, {Moulin}, {Naumann}, {Naumann-Godo}, {de Naurois}, {Nedbal},
  {Nekrassov}, {Nicholas}, {Niemiec}, {Nolan}, {Ohm}, {de O{\~n}a Wilhelmi},
  {Opitz}, {Ostrowski}, {Oya}, {Panter}, {Paz Arribas}, {Pedaletti},
  {Pelletier}, {Petrucci}, {Pita}, {P{\"u}hlhofer}, {Punch}, {Quirrenbach},
  {Raue}, {Rayner}, {Reimer}, {Reimer}, {Renaud}, {de Los Reyes}, {Rieger},
  {Ripken}, {Rob}, {Rosier-Lees}, {Rowell}, {Rudak}, {Rulten}, {Ruppel},
  {Ryde}, {Sahakian}, {Santangelo}, {Schlickeiser}, {Sch{\"o}ck}, {Schulz},
  {Schwanke}, {Schwarzburg}, {Schwemmer}, {Sikora}, {Skilton}, {Sol},
  {Spengler}, {Stawarz}, {Steenkamp}, {Stegmann}, {Stinzing}, {Stycz},
  {Sushch}, {Szostek}, {Tavernet}, {Terrier}, {Tluczykont}, {Tzioumis},
  {Valerius}, {van Eldik}, {Vasileiadis}, {Venter}, {Venter}, {Vialle},
  {Viana}, {Vincent}, {V{\"o}lk}, {Volpe}, {Vorobiov}, {Vorster}, {Wagner},
  {Ward}, {White}, {Wierzcholska}, {Zacharias}, {Zajczyk}, {Zdziarski}, {Zech},
  \& {Zechlin}}]{Abramowski2012A&A}
{H.~E.~S.~S. Collaboration}, {Abramowski}, A., {Acero}, F., {et~al.} 2012,
  \aap, 539, A149

\bibitem[{{HI4PI Collaboration} {et~al.}(2016){HI4PI Collaboration}, {Ben
  Bekhti}, {Fl{\"o}er}, {Keller}, {Kerp}, {Lenz}, {Winkel}, {Bailin},
  {Calabretta}, {Dedes}, {Ford}, {Gibson}, {Haud}, {Janowiecki}, {Kalberla},
  {Lockman}, {McClure-Griffiths}, {Murphy}, {Nakanishi}, {Pisano}, \&
  {Staveley-Smith}}]{HI4PI_2016}
{HI4PI Collaboration}, {Ben Bekhti}, N., {Fl{\"o}er}, L., {et~al.} 2016, \aap,
  594, A116

\bibitem[{{Hu} {et~al.}(2024{\natexlab{a}}){Hu}, {Wang}, {Xue}, {Peng}, \&
  {Wang}}]{Hu2024MNRAS}
{Hu}, H.-B., {Wang}, H.-Q., {Xue}, R., {Peng}, F.-K., \& {Wang}, Z.-R.
  2024{\natexlab{a}}, \mnras, 528, 7587

\bibitem[{{Hu} {et~al.}(2024{\natexlab{b}}){Hu}, {Kang}, {Cai}, {Wang}, {Su},
  \& {Xiao}}]{Hu2024ApJ}
{Hu}, W., {Kang}, J.-L., {Cai}, Z.-Y., {et~al.} 2024{\natexlab{b}}, \apj, 972,
  31

\bibitem[{{Hu} \& {Yan}(2021)}]{Hu2021MNRAS}
{Hu}, W. \& {Yan}, D. 2021, \mnras, 508, 4038

\bibitem[{{Hu} {et~al.}(2023){Hu}, {Yan}, \& {Hu}}]{Hu2023ApJ}
{Hu}, W., {Yan}, D.-H., \& {Hu}, Q.-L. 2023, \apj, 948, 82

\bibitem[{{Inoue} \& {Tanaka}(2016)}]{Inoue2016ApJ}
{Inoue}, Y. \& {Tanaka}, Y.~T. 2016, \apj, 828, 13

\bibitem[{{Jansen} {et~al.}(2001){Jansen}, {Lumb}, {Altieri}, {Clavel}, {Ehle},
  {Erd}, {Gabriel}, {Guainazzi}, {Gondoin}, {Much}, {Munoz}, {Santos},
  {Schartel}, {Texier}, \& {Vacanti}}]{Jansen2001A&A}
{Jansen}, F., {Lumb}, D., {Altieri}, B., {et~al.} 2001, \aap, 365, L1

\bibitem[{{Jones}(1968)}]{Jones1968}
{Jones}, F.~C. 1968, Physical Review, 167, 1159

\bibitem[{Kang \& Wang(2024)}]{Kang_2024}
Kang, J.-L. \& Wang, J.-X. 2024, JUSTC, 54, 2

\bibitem[{{Kirk} {et~al.}(1996){Kirk}, {Duffy}, \& {Gallant}}]{Kirk1996A&A}
{Kirk}, J.~G., {Duffy}, P., \& {Gallant}, Y.~A. 1996, \aap, 314, 1010

\bibitem[{{Konigl}(1981)}]{Konigl1981}
{Konigl}, A. 1981, \apj, 243, 700

\bibitem[{{Kouch} {et~al.}(2024){Kouch}, {Liodakis}, {Middei}, {Kim},
  {Tavecchio}, {Marscher}, {Marshall}, {Ehlert}, {Di Gesu}, {Jorstad}, {Agudo},
  {Madejski}, {Romani}, {Errando}, {Lindfors}, {Nilsson}, {Toppari}, {Potter},
  {Imazawa}, {Sasada}, {Fukazawa}, {Kawabata}, {Uemura}, {Mizuno}, {Nakaoka},
  {Akitaya}, {McCall}, {Jermak}, {Steele}, {Myserlis}, {Gurwell}, {Keating},
  {Rao}, {Kang}, {Lee}, {Kim}, {Cheong}, {Jeong}, {Angelakis}, {Kraus},
  {Aceituno}, {Bonnoli}, {Casanova}, {Escudero}, {Ag{\'\i}s-Gonz{\'a}lez},
  {Husillos}, {Morcuende}, {Otero-Santos}, {Sota}, {Bachev}, {Antonelli},
  {Bachetti}, {Baldini}, {Baumgartner}, {Bellazzini}, {Bianchi}, {Bongiorno},
  {Bonino}, {Brez}, {Bucciantini}, {Capitanio}, {Castellano}, {Cavazzuti},
  {Chen}, {Ciprini}, {Costa}, {De Rosa}, {Del Monte}, {Di Lalla}, {Di Marco},
  {Donnarumma}, {Doroshenko}, {Dov{\v{c}}iak}, {Enoto}, {Evangelista},
  {Fabiani}, {Ferrazzoli}, {Garcia}, {Gunji}, {Hayashida}, {Heyl}, {Iwakiri},
  {Kaaret}, {Karas}, {Kislat}, {Kitaguchi}, {Kolodziejczak}, {Krawczynski}, {La
  Monaca}, {Latronico}, {Maldera}, {Manfreda}, {Marin}, {Marinucci}, {Massaro},
  {Matt}, {Mitsuishi}, {Muleri}, {Negro}, {Ng}, {O'Dell}, {Omodei},
  {Oppedisano}, {Papitto}, {Pavlov}, {Peirson}, {Perri}, {Pesce-Rollins},
  {Petrucci}, {Pilia}, {Possenti}, {Poutanen}, {Puccetti}, {Ramsey}, {Rankin},
  {Ratheesh}, {Roberts}, {Sgr{\`o}}, {Slane}, {Soffitta}, {Spandre}, {Swartz},
  {Tamagawa}, {Taverna}, {Tawara}, {Tennant}, {Thomas}, {Tombesi}, {Trois},
  {Tsygankov}, {Turolla}, {Vink}, {Weisskopf}, {Wu}, {Xie}, \&
  {Zane}}]{Kouch2024A&A}
{Kouch}, P.~M., {Liodakis}, I., {Middei}, R., {et~al.} 2024, \aap, 689, A119

\bibitem[{{Lachowicz} {et~al.}(2009){Lachowicz}, {Gupta}, {Gaur}, \&
  {Wiita}}]{Lachowicz2009A&A}
{Lachowicz}, P., {Gupta}, A.~C., {Gaur}, H., \& {Wiita}, P.~J. 2009, \aap, 506,
  L17

\bibitem[{{Lewis} {et~al.}(2016){Lewis}, {Becker}, \& {Finke}}]{Lewis2016}
{Lewis}, T.~R., {Becker}, P.~A., \& {Finke}, J.~D. 2016, \apj, 824, 108

\bibitem[{{Lewis} {et~al.}(2018){Lewis}, {Finke}, \& {Becker}}]{Lewis2018ApJ}
{Lewis}, T.~R., {Finke}, J.~D., \& {Becker}, P.~A. 2018, \apj, 853, 6

\bibitem[{Madejski \& Sikora(2016)}]{Madejski2016annurev}
Madejski, G.~G. \& Sikora, M. 2016, Annual Review of Astronomy and
  Astrophysics, 54, 725

\bibitem[{{Madejski} {et~al.}(2016){Madejski}, {Nalewajko}, {Madsen}, {Chiang},
  {Balokovi{\'c}}, {Paneque}, {Furniss}, {Hayashida}, {Urry}, {Sikora},
  {Ajello}, {Blandford}, {Harrison}, {Sanchez}, {Giebels}, {Stern},
  {Alexander}, {Barret}, {Boggs}, {Christensen}, {Craig}, {Forster}, {Giommi},
  {Grefenstette}, {Hailey}, {Hornstrup}, {Kitaguchi}, {Koglin}, {Mao},
  {Miyasaka}, {Mori}, {Perri}, {Pivovaroff}, {Puccetti}, {Rana}, {Westergaard},
  {Zhang}, \& {Zoglauer}}]{Madejski2016ApJ}
{Madejski}, G.~M., {Nalewajko}, K., {Madsen}, K.~K., {et~al.} 2016, \apj, 831,
  142

\bibitem[{{Mangalam} \& {Wiita}(1993)}]{Mangalam1993ApJ}
{Mangalam}, A.~V. \& {Wiita}, P.~J. 1993, \apj, 406, 420

\bibitem[{{Marcowith} {et~al.}(2020){Marcowith}, {Ferrand}, {Grech}, {Meliani},
  {Plotnikov}, \& {Walder}}]{Marcowith2020LRCA}
{Marcowith}, A., {Ferrand}, G., {Grech}, M., {et~al.} 2020, Living Reviews in
  Computational Astrophysics, 6, 1

\bibitem[{{Marscher}(2014)}]{Marscher2014ApJ}
{Marscher}, A.~P. 2014, \apj, 780, 87

\bibitem[{{Massaro} {et~al.}(2008){Massaro}, {Tramacere}, {Cavaliere}, {Perri},
  \& {Giommi}}]{Massaro2008A&A}
{Massaro}, F., {Tramacere}, A., {Cavaliere}, A., {Perri}, M., \& {Giommi}, P.
  2008, \aap, 478, 395

\bibitem[{{McKinney} \& {Blandford}(2009)}]{McKinney2009MNRAS}
{McKinney}, J.~C. \& {Blandford}, R.~D. 2009, \mnras, 394, L126

\bibitem[{{Moderski} {et~al.}(2003){Moderski}, {Sikora}, \&
  {B{\l}a{\.z}ejowski}}]{Moderski2003A&A}
{Moderski}, R., {Sikora}, M., \& {B{\l}a{\.z}ejowski}, M. 2003, \aap, 406, 855

\bibitem[{{Netzer}(2015)}]{Netzer2015ARA&A}
{Netzer}, H. 2015, \araa, 53, 365

\bibitem[{{Ng} {et~al.}(2010){Ng}, {D{\'\i}az Trigo}, {Cadolle Bel}, \&
  {Migliari}}]{Ng_2010}
{Ng}, C., {D{\'\i}az Trigo}, M., {Cadolle Bel}, M., \& {Migliari}, S. 2010,
  \aap, 522, A96

\bibitem[{{Noel} {et~al.}(2022){Noel}, {Gaur}, {Gupta}, {Wierzcholska},
  {Ostrowski}, {Dhiman}, \& {Bhatta}}]{Noel2022ApJS}
{Noel}, A.~P., {Gaur}, H., {Gupta}, A.~C., {et~al.} 2022, \apjs, 262, 4

\bibitem[{{Pavana Gowtami} {et~al.}(2022){Pavana Gowtami}, {Gaur}, {Gupta},
  {Wiita}, {Liao}, \& {Ward}}]{Pavana2022MNRAS}
{Pavana Gowtami}, G.~S., {Gaur}, H., {Gupta}, A.~C., {et~al.} 2022, \mnras,
  511, 3101

\bibitem[{{Peceur} {et~al.}(2020){Peceur}, {Taylor}, \&
  {Kraan-Korteweg}}]{Peceur2020MNRAS}
{Peceur}, N.~W., {Taylor}, A.~R., \& {Kraan-Korteweg}, R.~C. 2020, \mnras, 495,
  2162

\bibitem[{{Pekeur} {et~al.}(2016){Pekeur}, {Taylor}, {Potter}, \&
  {Kraan-Korteweg}}]{Pekeur2016MNRAS}
{Pekeur}, N.~W., {Taylor}, A.~R., {Potter}, S.~B., \& {Kraan-Korteweg}, R.~C.
  2016, \mnras, 462, L80

\bibitem[{{Peterson} {et~al.}(1998){Peterson}, {Wanders}, {Horne}, {Collier},
  {Alexander}, {Kaspi}, \& {Maoz}}]{Peterson1998PASP}
{Peterson}, B.~M., {Wanders}, I., {Horne}, K., {et~al.} 1998, \pasp, 110, 660

\bibitem[{{Petropoulou} {et~al.}(2019){Petropoulou}, {Sironi}, {Spitkovsky}, \&
  {Giannios}}]{Petropoulou2019ApJ}
{Petropoulou}, M., {Sironi}, L., {Spitkovsky}, A., \& {Giannios}, D. 2019,
  \apj, 880, 37

\bibitem[{{Potter} \& {Cotter}(2013)}]{Potter2013MNRAS}
{Potter}, W.~J. \& {Cotter}, G. 2013, \mnras, 429, 1189

\bibitem[{{Protheroe} \& {Clay}(2004)}]{Protheroe2004PASA}
{Protheroe}, R.~J. \& {Clay}, R.~W. 2004, \pasa, 21, 1

\bibitem[{{Rieger} {et~al.}(2007){Rieger}, {Bosch-Ramon}, \&
  {Duffy}}]{Rieger2007Ap&SS}
{Rieger}, F.~M., {Bosch-Ramon}, V., \& {Duffy}, P. 2007, \apss, 309, 119

\bibitem[{{Sasaki} {et~al.}(2015){Sasaki}, {Asano}, \&
  {Terasawa}}]{Sasaki2015ApJ}
{Sasaki}, K., {Asano}, K., \& {Terasawa}, T. 2015, \apj, 814, 93

\bibitem[{{Schlickeiser}(1984)}]{Schlickeiser1984}
{Schlickeiser}, R. 1984, \aap, 136, 227

\bibitem[{{Schwartz} {et~al.}(1979){Schwartz}, {Doxsey}, {Griffiths},
  {Johnston}, \& {Schwarz}}]{Schwartz1979ApJ}
{Schwartz}, D.~A., {Doxsey}, R.~E., {Griffiths}, R.~E., {Johnston}, M.~D., \&
  {Schwarz}, J. 1979, \apjl, 229, L53

\bibitem[{{Sironi} \& {Spitkovsky}(2009)}]{Sironi2009ApJ}
{Sironi}, L. \& {Spitkovsky}, A. 2009, \apjl, 707, L92

\bibitem[{{Sironi} {et~al.}(2013){Sironi}, {Spitkovsky}, \&
  {Arons}}]{Sironi2013ApJ}
{Sironi}, L., {Spitkovsky}, A., \& {Arons}, J. 2013, \apj, 771, 54

\bibitem[{{Stathopoulos} {et~al.}(2024){Stathopoulos}, {Petropoulou},
  {Vasilopoulos}, \& {Mastichiadis}}]{Stathopoulos2024A}
{Stathopoulos}, S.~I., {Petropoulou}, M., {Vasilopoulos}, G., \&
  {Mastichiadis}, A. 2024, \aap, 683, A225

\bibitem[{{Stawarz} \& {Petrosian}(2008)}]{Stawarz2008}
{Stawarz}, {\L}. \& {Petrosian}, V. 2008, \apj, 681, 1725

\bibitem[{{Str{\"u}der} {et~al.}(2001){Str{\"u}der}, {Briel}, {Dennerl},
  {Hartmann}, {Kendziorra}, {Meidinger}, {Pfeffermann}, {Reppin}, {Aschenbach},
  {Bornemann}, {Br{\"a}uninger}, {Burkert}, {Elender}, {Freyberg}, {Haberl},
  {Hartner}, {Heuschmann}, {Hippmann}, {Kastelic}, {Kemmer}, {Kettenring},
  {Kink}, {Krause}, {M{\"u}ller}, {Oppitz}, {Pietsch}, {Popp}, {Predehl},
  {Read}, {Stephan}, {St{\"o}tter}, {Tr{\"u}mper}, {Holl}, {Kemmer}, {Soltau},
  {St{\"o}tter}, {Weber}, {Weichert}, {von Zanthier}, {Carathanassis}, {Lutz},
  {Richter}, {Solc}, {B{\"o}ttcher}, {Kuster}, {Staubert}, {Abbey}, {Holland},
  {Turner}, {Balasini}, {Bignami}, {La Palombara}, {Villa}, {Buttler},
  {Gianini}, {Lain{\'e}}, {Lumb}, \& {Dhez}}]{Struder2001A&A}
{Str{\"u}der}, L., {Briel}, U., {Dennerl}, K., {et~al.} 2001, \aap, 365, L18

\bibitem[{{Summerlin} \& {Baring}(2012)}]{Summerlin2012ApJ}
{Summerlin}, E.~J. \& {Baring}, M.~G. 2012, \apj, 745, 63

\bibitem[{{Tan} {et~al.}(2024){Tan}, {Liu}, \& {B{\"o}ttcher}}]{Tan2024MNRAS}
{Tan}, H.-B., {Liu}, R.-Y., \& {B{\"o}ttcher}, M. 2024, \mnras, 529, 903

\bibitem[{{Thiersen} {et~al.}(2022){Thiersen}, {Zacharias}, \&
  {B{\"o}ttcher}}]{Thiersen2022ApJ}
{Thiersen}, H., {Zacharias}, M., \& {B{\"o}ttcher}, M. 2022, \apj, 925, 177

\bibitem[{{Urry} \& {Padovani}(1995)}]{Urry1995PASP}
{Urry}, C.~M. \& {Padovani}, P. 1995, \pasp, 107, 803

\bibitem[{{Ushio} {et~al.}(2010){Ushio}, {Stawarz}, {Takahashi}, {Paneque},
  {Madejski}, {Hayashida}, {Kataoka}, {Tanaka}, {Tanaka}, \&
  {Ostrowski}}]{Ushio2010}
{Ushio}, M., {Stawarz}, {\L}., {Takahashi}, T., {et~al.} 2010, \apj, 724, 1509

\bibitem[{{Uttley} {et~al.}(2014){Uttley}, {Cackett}, {Fabian}, {Kara}, \&
  {Wilkins}}]{Uttley2014AARv}
{Uttley}, P., {Cackett}, E.~M., {Fabian}, A.~C., {Kara}, E., \& {Wilkins},
  D.~R. 2014, \aapr, 22, 72

\bibitem[{{VanderPlas}(2018)}]{VanderPlas2018ApJS}
{VanderPlas}, J.~T. 2018, \apjs, 236, 16

\bibitem[{{Vaughan} \& {Nowak}(1997)}]{Vaughan1997ApJ}
{Vaughan}, B.~A. \& {Nowak}, M.~A. 1997, \apjl, 474, L43

\bibitem[{{Vaughan}(2010)}]{Vaughan2010MNRAS}
{Vaughan}, S. 2010, \mnras, 402, 307

\bibitem[{{Vuillaume} {et~al.}(2018){Vuillaume}, {Henri}, \&
  {Petrucci}}]{Vuillaume2018A&A}
{Vuillaume}, T., {Henri}, G., \& {Petrucci}, P.~O. 2018, \aap, 620, A41

\bibitem[{Wang {et~al.}(2022)Wang, Liu, Petropoulou, Oikonomou, Xue, \&
  Wang}]{WangPhysRevD}
Wang, Z.-R., Liu, R.-Y., Petropoulou, M., {et~al.} 2022, Phys. Rev. D, 105,
  023005

\bibitem[{{Wilkins}(2019)}]{Wilkins2019MNRAS}
{Wilkins}, D.~R. 2019, \mnras, 489, 1957

\bibitem[{{Xu} {et~al.}(2023){Xu}, {Hu}, {Chen}, {Jiang}, \&
  {Alexeeva}}]{Xu2023ApJS}
{Xu}, J., {Hu}, S., {Chen}, X., {Jiang}, Y., \& {Alexeeva}, S. 2023, \apjs,
  268, 54

\bibitem[{{Zamaninasab} {et~al.}(2014){Zamaninasab}, {Clausen-Brown},
  {Savolainen}, \& {Tchekhovskoy}}]{Zamaninasab2014Natur}
{Zamaninasab}, M., {Clausen-Brown}, E., {Savolainen}, T., \& {Tchekhovskoy}, A.
  2014, \nat, 510, 126

\bibitem[{{Zhang} \& {Giannios}(2021)}]{Zhang2021MNRAS}
{Zhang}, H. \& {Giannios}, D. 2021, \mnras, 502, 1145

\bibitem[{{Zhang} {et~al.}(2022){Zhang}, {Yan}, \& {Zhang}}]{Zhang2022ApJ}
{Zhang}, H., {Yan}, D., \& {Zhang}, L. 2022, \apj, 930, 157

\bibitem[{{Zhang} {et~al.}(2023){Zhang}, {Yan}, \& {Zhang}}]{Zhang2023ApJ}
{Zhang}, H., {Yan}, D., \& {Zhang}, L. 2023, \apj, 944, 103

\bibitem[{{Zhang} {et~al.}(2021{\natexlab{a}}){Zhang}, {Yan}, {Zhang}, {Yang},
  \& {Zhang}}]{Zhanghy2021ApJ}
{Zhang}, H., {Yan}, D., {Zhang}, P., {Yang}, S., \& {Zhang}, L.
  2021{\natexlab{a}}, \apj, 919, 58

\bibitem[{{Zhang} {et~al.}(2006){Zhang}, {Treves}, {Maraschi}, {Bai}, \&
  {Liu}}]{Zhang2006ApJ}
{Zhang}, Y.~H., {Treves}, A., {Maraschi}, L., {Bai}, J.~M., \& {Liu}, F.~K.
  2006, \apj, 637, 699

\bibitem[{{Zhang} {et~al.}(2021{\natexlab{b}}){Zhang}, {Gupta}, {Gaur},
  {Wiita}, {An}, {Lu}, {Fan}, \& {Xu}}]{Zhongli2021ApJ}
{Zhang}, Z., {Gupta}, A.~C., {Gaur}, H., {et~al.} 2021{\natexlab{b}}, \apj,
  909, 103

\end{thebibliography}

\begin{appendix} 
\onecolumn

\section{Comparison between observational and theoretical light curves}\label{sec:append1}
\begin{figure*}[h!]
   \centering
  \includegraphics[width=0.8\textwidth]{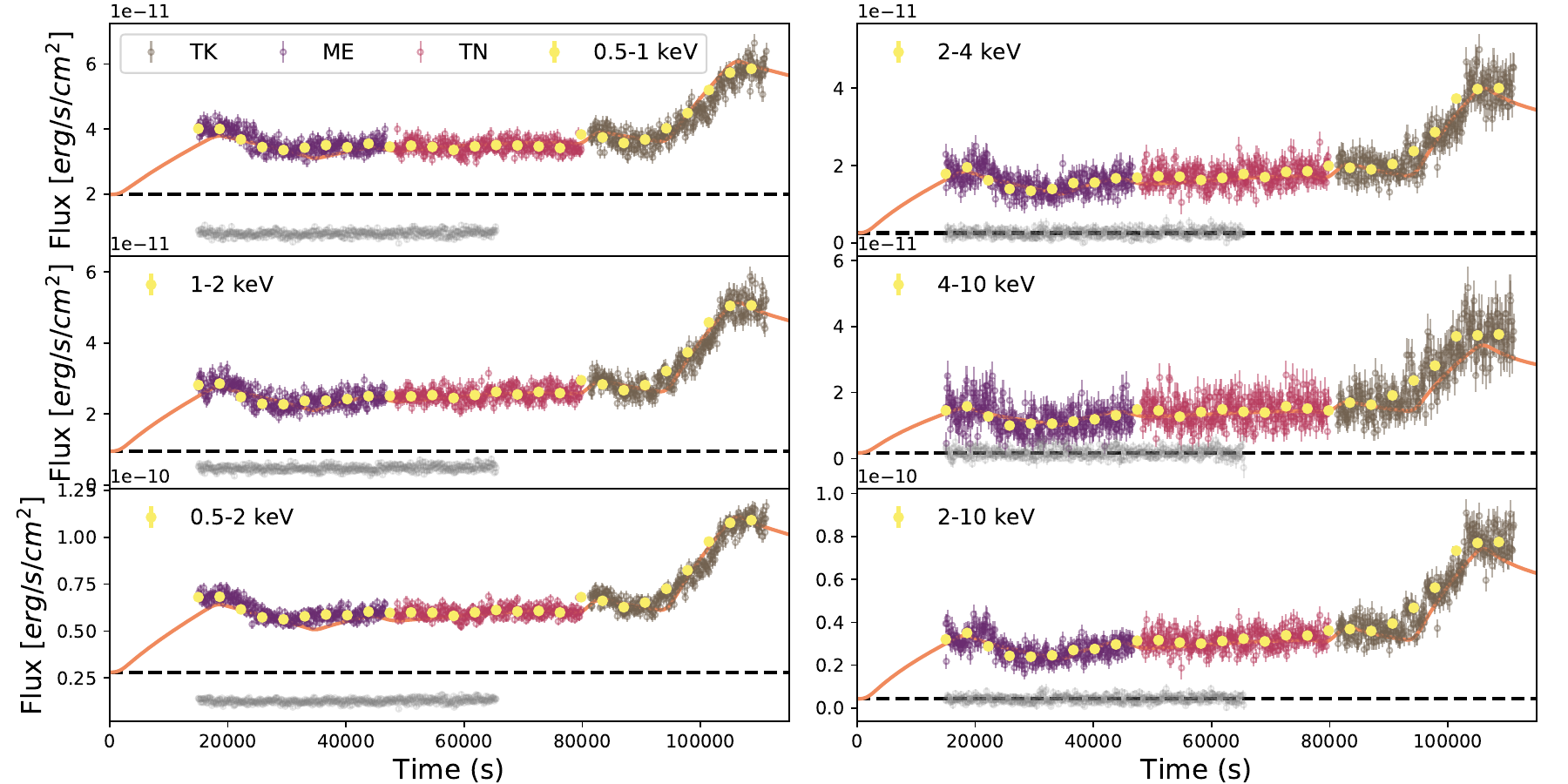}
   \caption{\textit{XMM-Newton} EPIC-on LCs from ObsID 0124930501. TK, ME, and TN represent the data observed with Thick, Median, and Thin filters, respectively. The  yellow data points denote the 1 hr binned LCs.
    The grey circles denote the lowest X-ray flux observed in ObsID 0411780701.  
    The dashed lines represent the flux level from the quasi-stationary zone. The orange solid lines represent the theoretical LCs reproduced by our two-zone model.}
              \label{figs:lc2}%
 \end{figure*}

\begin{figure*}[h!]
   \centering
  \includegraphics[width=0.8\textwidth]{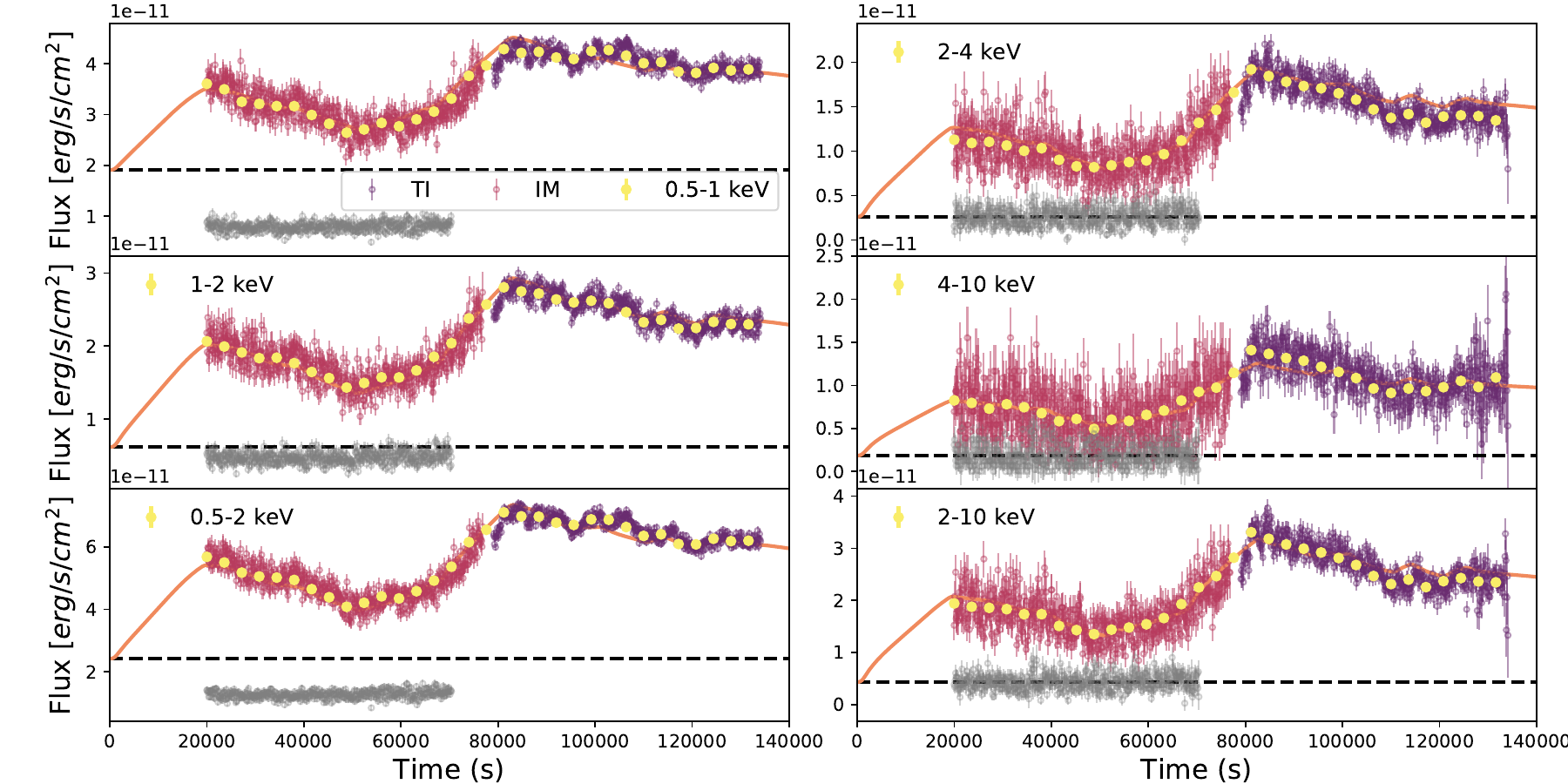}
   \caption{\textit{XMM-Newton} EPIC-pn LCs from ObsID 0124930601. TI and IM represent the data in Timing and Imaging science modes, respectively. The yellow data points denote the 1 hr binned LCs.
    The grey circles denote the lowest X-ray flux observed in ObsID 0411780701.  
    The dashed lines represent the flux level from the quasi-stationary zone. The orange solid lines represent the theoretical LCs reproduced by our two-zone model.}
              \label{figs:lc3}%
\end{figure*}

The observational and theoretical light curves (LCs) in various sub-bands for ObsID 0124930501 and 0124930601 are shown in Figs. \ref{figs:lc2} and \ref{figs:lc3}, respectively.

\section{Power spectral density and lag-frequency spectra}\label{sec:append2}

\begin{figure*}[h!]
   \centering
  \includegraphics[width=0.7\textwidth]{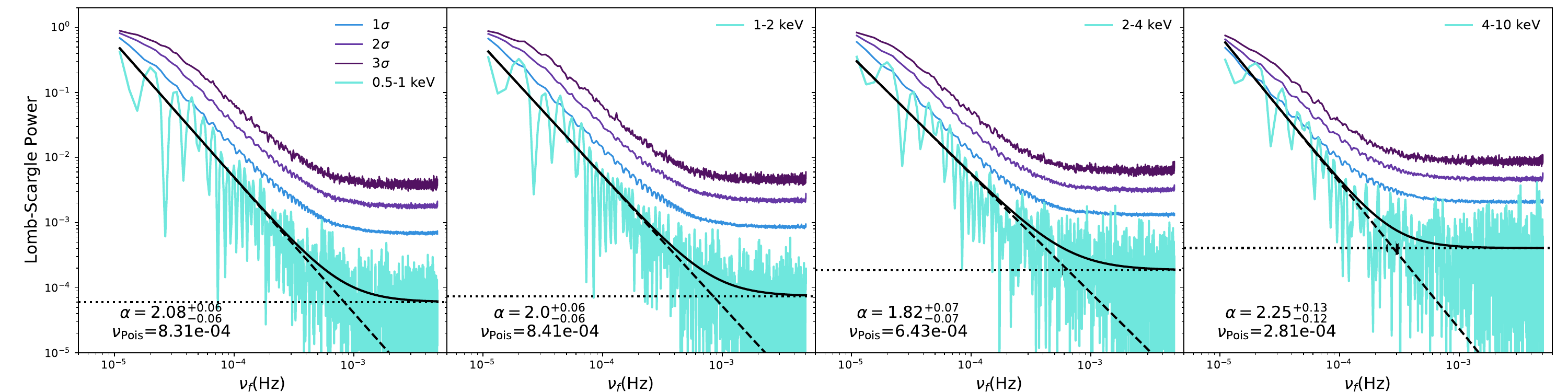}  
  \includegraphics[width=0.7\textwidth]{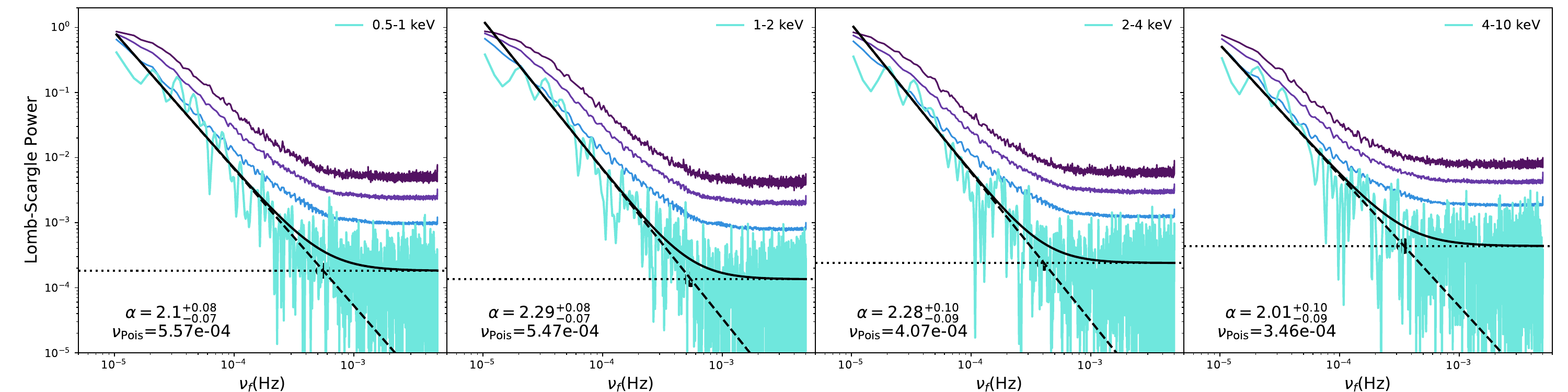}  
  \includegraphics[width=0.7\textwidth]{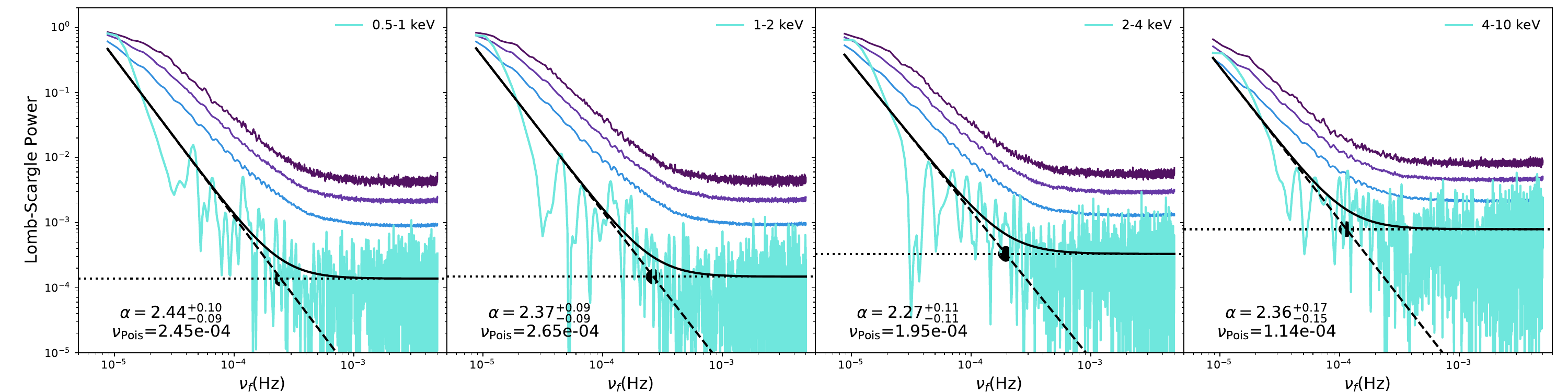}
   \caption{PSDs of PKS 2155-304 in the four sub-band observed from ObsIDs 0124930301 (top), 0124930501 (middle), and 0124930601 (bottom). The black solid circles mark the frequency threshold at which white noise starts to dominate the PSD
   and the dotted horizontal lines mark the white noise level for each band. 
  1$\sigma$, 2$\sigma$, and 3$\sigma$ significance levels are calculated based on the simulation of 5000 LCs with the Gaussian process (GP) method.
The resulting power is a dimensionless quantity that lies in the range $0 < P < 1$.}
              \label{figs:psd}%
    \end{figure*}

\begin{figure*}[h!]
   \centering
  \includegraphics[width=0.75\textwidth]{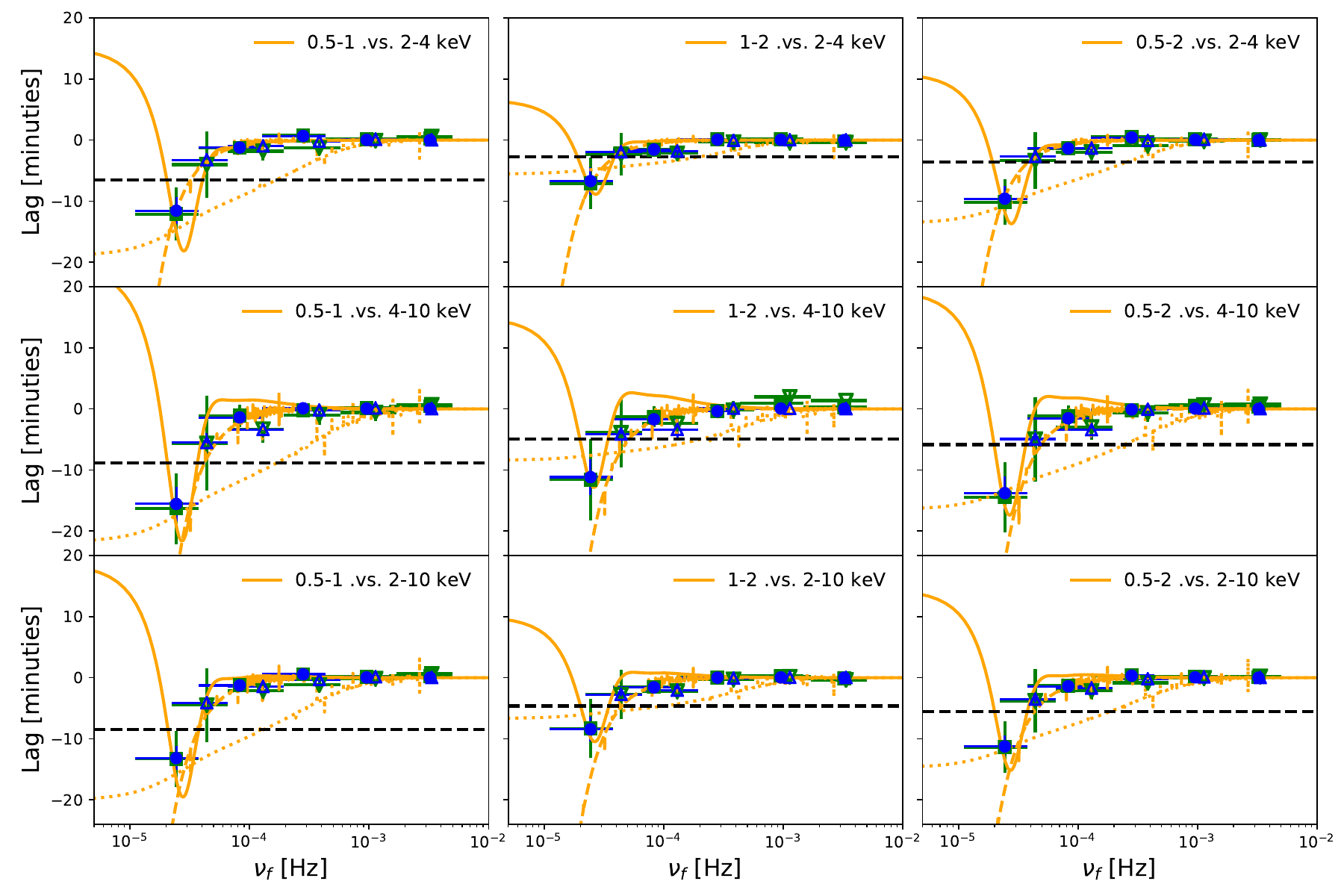}
   \caption{
    Fourier lag-frequency spectra measured in ObsID 0124930301. 
   The measured lag-frequency spectra $\tau_1$ and $\tau_2$ are denoted by the closed and opened symbols, respectively (see text for details).
   The green squares and inverted triangles are the lag-frequency spectra calculated by the linearly interpolated LCs,
   while the blue points and triangles are the lag-frequency spectra calculated by the simulated LCs with GP. 
   The orange dotted and dashed lines respectively represent the theoretical time-lag spectra expected from Case A and B in the frame of the classical one-zone synchrotron self-Compton (SSC) model.   
   The horizontal dashed lines denote the time lags approximately calculated by the analytical method for Case A.
   The orange solid lines are the theoretical predictions from our proposed two-zone SSC model.
   The detail discussion is presented in Sect. \ref{sec:discu}. 
} \label{figs:ftlag1}%
 \end{figure*}

\begin{figure*}
   \centering
  \includegraphics[width=0.75\textwidth]{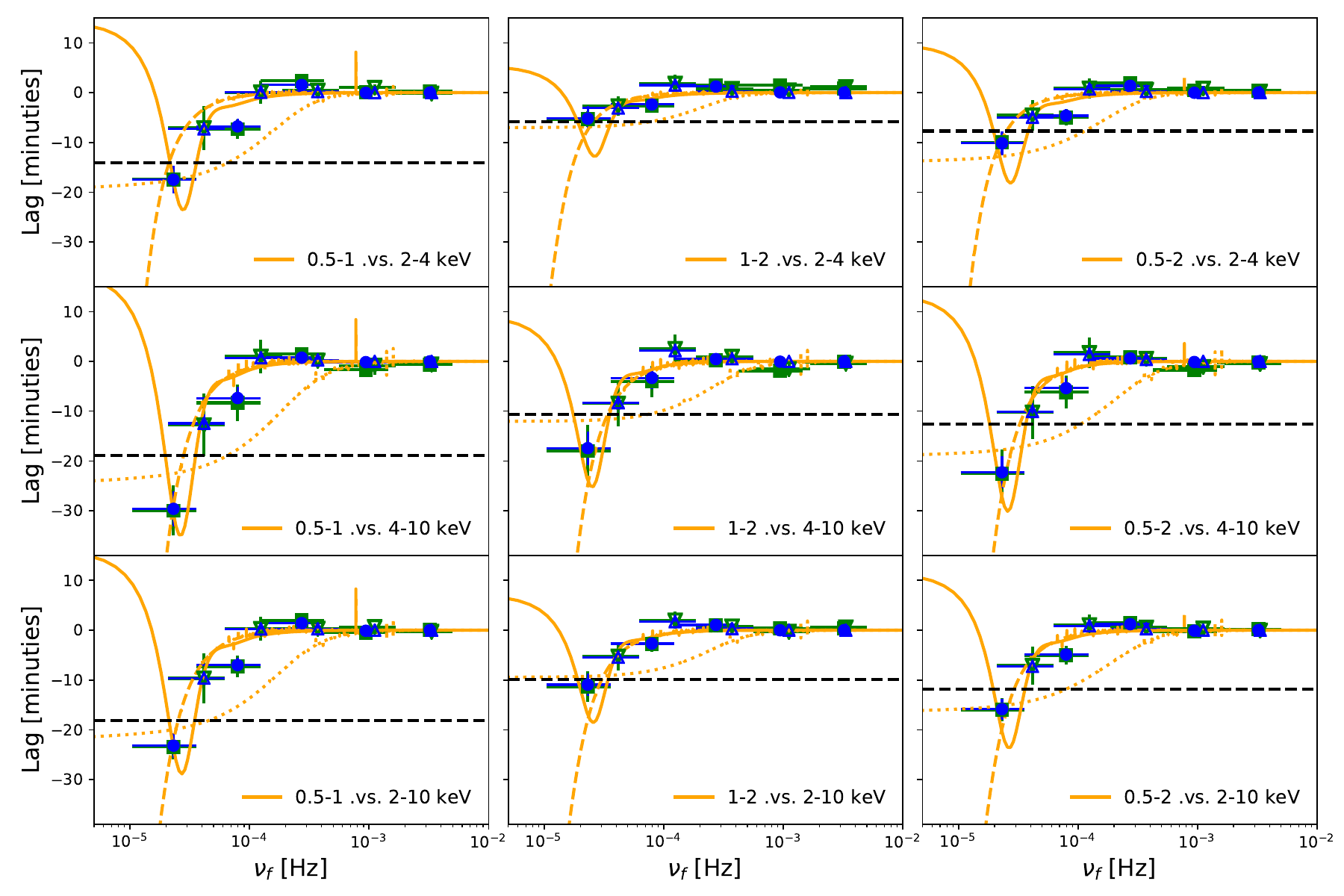}
   \caption{Fourier lag-frequency spectra measured in ObsID 0124930501. Same as Fig.\ref{figs:ftlag1}.
   }
              \label{figs:ftlag2}%
    \end{figure*}

\begin{figure*}
   \centering
  \includegraphics[width=0.75\textwidth]{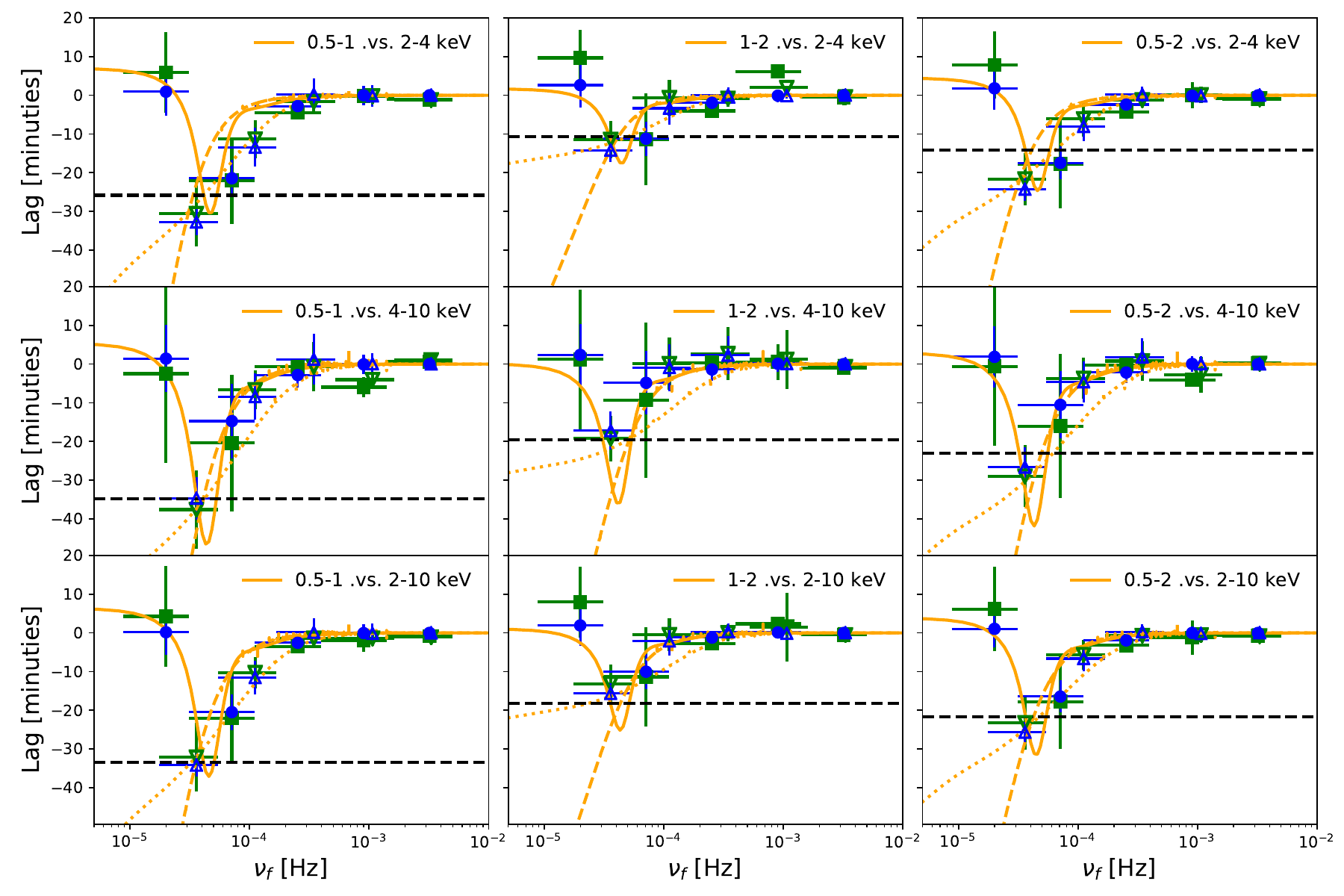}
   \caption{Fourier lag-frequency spectra measured in ObsID 0124930601. Same as Fig.\ref{figs:ftlag1}.
 }
    \label{figs:ftlag3}%
 \end{figure*}

 The measured power spectral densities (PSDs) in various sub-bands for three observations are shown in Fig. \ref{figs:psd}
 and the lag-frequency spectra are presented in Figs. \ref{figs:ftlag1} to \ref{figs:ftlag3}.

\end{appendix}

\end{document}